\documentstyle[12pt]{article}   
\textwidth160mm\hoffset-12mm  \title{A quantum statistical mechanics model of a
three dimensional linear rigid rotator in a bath of oscillators:
 IV- steady state dielectric properties induced by a.c and d.c field
coupling.}   
 \author{ J. T. Titantah\addtocounter{footnote}{+1}\footnote{Present adress: \small \it D\'epartement de Physique, Universit\'e Libre de Bruxelles, Brussels (Belgium)}\mbox{ } and M. N. Hounkonnou} \date{ Unit\'e de Recherche en
Physique Th\'eorique,\\ Institut de Math\'ematiques et de Sciences Physiques,\\
B.P.: 613 Porto-Novo - Republic of Benin.}      
 
\begin{document} \def\t{(Bt)} \addtolength{\baselineskip}{5mm}           
\maketitle

\begin{abstract} The long time effect of a radio frequency (rf) a.c. field
superimposed on a d.c. field on the electrical susceptibility and the Kerr
optical functions of polarisable fluids in inert solvent is analysed. The
results obtained for the classical Brownian limit, valid for  dense solvent
media, well reproduce  classical results published in the literature with
excellent precisions in inertia, density and temperature dependences.
The low density limit yields absorption-dispersion lines whose widths and shifts are
density, inertia and temperature dependent. While the low density and/or large
inertia susceptibility is explicitly written out as a continued fraction got by
solving an infinite hierarchy of differential coupled equations, that of the
Kerr effect is given in the form of successive convergents of the solutions of
an infinite hierarchy of differential difference triplets. The polarisation/a.c.\ field phase difference is analysed. The effects of the
constant field strength and the a.c. field frequency on the Kerr function are
explored. In this paper, which will be named paper IV, the derivation of some
quoted equations will intensionally be left out as they exist in paper III
(\textit{J. Phys. A}, 30:6347, 1997) of
which this work is its logical continuation.\end{abstract}     
 \newpage
\section{Introduction} In recent works \cite{navhoun95I,tit96II,tit96III}, we
derived a master equation describing the evolution of a system of linear rigid 
rotators in a bath of non interacting harmonic oscillators. Due to the fact that
experimentally observed spectra are accounted for, strongly, by molecular
rotational motions, we consider the rotational degrees of freedom of the
rotator by defining an orientation operator $\hat{\bf u}=\hat{\bf\mu}/\mu$,
where $\hat{\bf\mu}$ is the rotator permanent dipole moment of magnitude $\mu$.
The master equation for the statistical orientation probability density operator
$\hat{\rho}_S(t)$ associated with the motion of the rotator in interaction with
the bath and an external driving field is \cite{navhoun95I,tit96II}
 \begin {equation}
 \frac{\partial \hat{\rho}_S (t)}{\partial t}+ \frac{i}{\hbar}\Bigl
[\hat{H}_S,\hat{\rho}_S (t)\Bigr]
+\hat{K}\hat{\rho}_S(t)=-\frac{i}{\hbar}\Bigl[\hat{H}_E,\hat{\rho}_S
(t)\Bigr],\label{eq:master}\end{equation} where $\hat{H}_S$ is the rotator
rotational kinetic energy operator, whose eigen-values are
$E_l=(\hbar^2/2I)l(l+1) $ with $l=0,1,2,...$ 

The collision term is written as 
\begin {eqnarray} \hat{K} \hat{\rho}_S(t)&=&\frac{\zeta}{I} \sum_{l=1} ^{\infty}
l  \Bigl\{A_l ^* \hat{\bf{u}}.\hat{\bf{u}}_l^- \hat{\rho} _ S (t) - A_l  
\hat{\bf{u}}.\hat{\rho}_S (t )\hat{\bf{u}}_l ^+ + B_l\hat{\bf{u}}.
\hat{\bf{u}}_l ^+ \hat{\rho} _S (t)  - B_l ^*  \hat{\bf{u}}.\hat {\rho}_S
(t)\hat{\bf{u}}_l ^ - \nonumber \\ & & - A_l ^*  \hat{\bf {u}}_l ^-.\hat{\rho}_S
(t ) \hat{\bf{u}} + A_l \hat{ \rho}_S(t) \hat{\bf{u}}_l ^+.\hat{\bf{u}} + B_l ^*
\hat{ \rho}_S (t) \hat {\bf{u}}_l ^-.\hat{\bf{u}} -  B_l \hat{\bf{u}}_l ^+ 
\hat{\rho}_S (t).\hat{\bf{u}} \Bigr\},  \end{eqnarray} where \begin{equation}
A_l =\frac{\omega_D ^2}{\omega_D ^2 + \omega_l ^2}\Bigl[1+N(\omega_l) +
i\Bigl(\kappa(x_l, x_D) - \frac{ \omega_l}{ 2\omega_D} \Bigr)\Bigr], \end
{equation}  \begin{equation} B_l =\frac{\omega_D ^2}{\omega_D ^2 + \omega_l
^2}\Bigl[N(\omega_l) +i\Bigl(\kappa(x_l, x_D)+ \frac{ \omega_l}{ 2\omega_D}
\Bigr)\Bigr], \end {equation} with \begin {equation} \kappa(x_l, x_D)= - \Bigl[
\frac{1}{x_D}+2 \sum_{n=1}^\infty \frac {x_l^2-2\pi x_Dn}{(x_l+x_D)(x_l^2+4\pi^2
n^2)} \Bigr]
 \end {equation} and  \begin {equation} x_D=\beta \hbar \omega_D, \mbox{ }
x_l=\beta \hbar \omega_l,\mbox{ }  \beta=1/(k_B T), \mbox{ } n =1,2,3,...
 \end {equation}  $\omega_D$ is the characteristic Debye frequency, $k_B$ the
Boltzmann constant, $T$ the absolute temperature  and $N(\omega_l)$ the
occupation number of the rotator quantum level $l$ (bosonic). $A_l^*$ and $B_l^*$ are the
complex conjugates of $A_l$ and $B_l$, respectively. $\zeta$ is the friction
coefficient characterising the effects of the bath oscillator concentration  on
the rotator dynamics, $I$ is the rotator moment of inertia.
 
 We use the spherical harmonic expansion of the unit vector operator $\hat
{\bf{u}}$ as \cite{navhoun95I,tit96II} \begin{equation}
 \hat {\bf{u}}(t)= \sum_{l=1}^ {\infty}( \hat {\bf{u}}_l ^ +  +  \hat
{\bf{u}}_l^- (t) ), \end{equation}  where  \begin{equation} \hat{u}_{lx}^+
(t)=\frac{1}{2} \sum_{m=-l}^ l  \mid l,m> [< l-1, m+1 \mid A(l,m) - < l-1, m-1
\mid B(l,m)], \end{equation}  \begin{equation} \hat{u}_{ly}^+ (t)=\frac{1}{2i}
\sum_{m=-l}^ l   \mid l,m > [< l-1, m+1 \mid A(l,m) + < l-1, m-1 \mid B(l,m)],
\end{equation}  \begin{equation} \hat{u}_{lz}^+ (t)= \sum_{m=-l}^ l  \mid l,m > 
< l-1, m \mid C(l,m)   \end{equation}
 and  \begin{equation} \hat {\bf{u}}_l ^- (t)=(\hat{\bf{u}}_{l}^+ (t))^{\dag}.
\end {equation}  \begin{equation} A(l, m) =  \sqrt{
\frac{(l-m)(l-m-1)}{(2l-1)(2l+1)}},\end{equation} \begin{equation} B(l, m) =
\sqrt{ \frac{(l+m)(l+m-1)}{(2l-1)(2l+1)}}, \end {equation}  \begin{equation}
C(l, m) =  \sqrt{\frac{(l-m)(l+m)}{(2l-1)(2l+1)}}. \end {equation}

The a.c.-d.c. field coupling term is \begin{eqnarray}\hat{H}_E(t) =
\left\{\begin{array}{ll} 0&\mbox{if  $t\leq0$}\\ - \mu \Bigl(E_c+E_a \cos \omega
t\Bigr) \cos \hat{ \beta}- \frac{\alpha_{\parallel}-
\alpha_\perp}{2}\Bigl(E_c+E_a\cos \omega t\Bigr)^2 \cos^2{\hat{\beta}}-\\- \frac{
\alpha_\perp}{2}\Bigl( E_c+E_a\cos\omega t\Bigr)^2 \hat{I} &\mbox{if  $t>0$}. 
\end{array}  \right. \end{eqnarray} $\alpha_{\parallel}$ and $\alpha_\perp$ are,
respectively, the rotator
 polarisability tensor components parallel and perpendicular  to the molecular
principal axis. We have assumed that the electric  fields are applied along the
$z$-axis of the laboratory frame. $\hat{\beta}$ is the angle between the applied field and the dipolar axis. With this Hamiltonian, the initial condition corresponds to equilibrium under free rotations, $\hat{\rho}(t=0)= \hat{\rho}_S^{eq}=\exp(-{\hat{H}_S \over k_BT})/Z$ where $Z$ is the rotator one particle partition function.

In a recent work \cite{tit96III}, we used the Hounkonnou-Navez master equation
\cite{navhoun95I} to verify the dielectric properties of a system of polar
rotators in interaction with a constant electric field of strength $E_c$. The
effects of inertia and bath concentration were intensively explored. Using the
rotational Smoluchowski equation \cite {morita78},  Morita et al.
\cite{morita79} and Matsumoto et al. \cite{mawayo70}  presented  studies of this problem but the former laid much interest on the effect of the applied field on the Kerr effect relaxation that results from the sudden application of the d.c. field. By averaging the Langevin equation, Coffey \cite{cof90,coff90,cof91}
tackled the problem emphasising on the effect of inertia but his analysis was
limited to the electrical susceptibility. 

In the present work, we consider the effect of coupling a constant d.c. field
with a radio frequency a.c. field. In the course of this work, we adopt the
notations of paper III  \cite{tit96III} so that we can
directly exploit existing results therein. The polarisation and the Kerr
functions can be calculated using the Hounkonnou-Titantah (HT) quantum relations
\cite{tit96III}\begin{equation} P(t)={2\mu\over
3}\sum_{l=0}^\infty(l+1){e^{-\beta E_l} \over Z}Re\sigma_{l,l+1}(t)
\label{eq:polar}  \end{equation}  \begin{equation}\Phi(t)={2\over
15}\sum_{l=0}^\infty {e^{-\beta E_l}\over Z} {(l+1)\over(2l+3)}\Biggl\{{l(2l+1)
\over (2l-1)}\varphi_{l,l}(t)+3(l+2)Re\eta_{l,l+2}(t)\Biggr\}\label{eq:kerr}. 
\end{equation}
 where $Z$ is the one particle free rotator canonical partition function and
$Re$ denotes the real part. The reduced HT equations for the matrix
elements $\sigma_{l,l+1}(t)$, $\varphi_{l,l}(t)$, and
$\eta_{l,l+2}(t)$ \cite{tit96III} are: 
 
 i) the reduced HT1: \begin{eqnarray} &
&\Bigl({\partial\over\partial t}-{i\hbar\over I}(l+1)\Bigr)\sigma_{l,l+1}(t)
+B\Bigl[\Bigl\{\Bigl(A_l^*l^2+B_{l+1}(l+1)^2\Bigr){1\over
2l+1}+(A_{l+1}(l+1)^2\nonumber\\& &+B_{l+2}^*(l+2)^2){1\over
2l+3}\Bigl\}\sigma_{l,l+1}(t)-e^{\beta\bigl(E_l- E_{l-1}\bigr)}{l\over
2l+1}\Bigl[B_ll\nonumber\\& &+
(l+1)B_{l+1}^*\Bigr]\sigma_{l-1,l}(t)(1-\delta_{l0})-
 e^{-\beta\bigl( E_{l+1}-E_l\bigr)}{l+2\over
2l+3}\Bigl[A_{l+1}^*(l+1)\nonumber\\& &+A_{l+2}(l+2)\Bigr]\sigma_{l+1,l+2}(t)-
{l+1\over(2l+1)(2l+3)}\Bigl[A_{l+1}^*(l+1)\nonumber\\&
&+B_{l+1}^*(+1)\Bigr]\sigma_ {l,l+1}^*(t)\Biggr] =-i{\mu E(t)\over
\hbar}\Biggl(1-e^{-\beta\bigl(E_{l+1}-E_l\bigr)}\Biggr);\label{eq:sigma1}\end{eqnarray}

ii) the reduced HT2:\begin {eqnarray} & &\frac{\partial}{\partial
t}\varphi_{l,l}(t)+2BRe\Biggl\{{\bigl(A_ll^2+B_{l+1}(l+1)^2\bigr)\over
2l+1}\varphi_{l,l}(t)-A_{l+1}(l+1){(l+2)(2l-1)\over (2l+1)^2}\nonumber\\ &
&\times e^{-\beta\bigl(E_{l+1}-E_l\bigr)}\varphi_{l+1,l+1}(t)-B_ll{(l-1)(2l+3)
\over(2l+1)^2}e^{\beta\bigl(E_l-E_{l-1}\bigr)}\varphi_{l-1,l-1}(t)(1-\delta_{l0})\nonumber\\
& &
 -3{(B_ll+A_{l+1}(l+1))\over
(2l+1)^2}e^{\beta\bigl(E_l-E_{l-1}\bigr)}\eta_{l-1,l+1}(t)(1-\delta_{l0})\Biggr\}
={\mu E(t) \over \hbar}\nonumber\\& &\times \Biggl({2l-1 \over
2l+1}{\it{Im}}\sigma_{l,l+1}(t)-e^{\beta\bigl(E_l-E_{l-1}\bigr)}{2l+3\over
2l+1}{\it{Im}}\sigma_{l-1,l}(t)(1-\delta_{l,0})\Biggr)\label{eq:phi1}
\end{eqnarray} and

iii) the reduced HT3: \begin {eqnarray} & &\Bigl[\frac{\partial}{\partial t}-
{i\hbar\over I}(2l+3)\Bigl]\eta_
{l,l+2}(t)+B\Biggl[\Bigl\{\Bigl[A_l^*l^2+B_{l+1}(l+1)^2\Bigr]{1\over2l+1}
+\Bigl[A_{l+2}(l+2)^2\nonumber\\& &+ B_{l+3}^*(l+3)^2\Bigr]{1\over
2l+5}\Bigl\}\eta_{l,l+2}(t)-{l\over 2l+1}e^{\beta(E_l-E_{l-1})} \nonumber\\&
&\times \Bigl[B_l^*l+B_{l+2}(l+2)\Bigr]\eta_{l-1,l+1}(t)(1-\delta_{l0})-
e^{-\beta\bigl( E_{l+1}-E_l\bigr)}{l+3\over 2l+5}\nonumber\\& &\times
\Bigl[A_{l+1}^*(l+1)+A_{l+3}(l+3)\Bigr]\eta_{l+1,l+3}(t) -{2\over
(2l+1)(2l+5)}\nonumber\\& &\times \Bigl[A_{l+1}^*(l+1)+B_{l+2}(+2)\Bigr]
\varphi_{l+1,l+1}(t) \Biggr]=i{\mu E(t)\over \hbar}\nonumber\\& &\times 
\Biggl(e^{-\beta
\bigl(E_{l+1}-E_l\bigr)}\sigma_{l+1,l+2}(t)-\sigma_{l,l+1}(t)\Biggr) -i{\Delta
\alpha E(t)^2\over 
2\hbar}\Biggl(1-e^{-\beta\bigl(E_{l+2}-E_l\bigr)}\Biggr),\label{eq:eta1}\end{eqnarray}
where $B=\zeta/I$. The initial conditions on these matrix elements are 

$\sigma_{l,l+1}(t=0)=\varphi_{l,l}(t=0)=\eta_{l,l+2}(t=0)=0$. 

\section{On the electrical susceptibility and the Kerr functions} In this
section, the calculations of the electrical susceptibility and the Kerr functions
are done in two different physical limits: 1) the classical
Brownian limit and 2) the rotating wave approximation (RWA). In each case, we
analyse the long time effect, that is, we consider times very long compared to
the period of collision $\tau=1/B$, the Debye relaxation time
$\tau_D=\zeta/(2Ik_BT)$ and the mean thermal angular time
$\Big(I/k_BT\Big)^{0.5}.$ 

\subsection{The classical Brownian limit} This limit
is characterised by slow moving rotators entering into instantaneous collisions
with the bath of fast moving oscillators. Inertial effects are very
important for understanding line shapes. With the aid of the Fokker-Planck-Kramer
(FPK) equation \cite{connel80,linden91,morita82,hounnavronv94,navhoun94},
Hounkonnou et al. \cite{hounnavronv94,hounronvhaz91,houn91,hounronv92} presented
the steady state analysis of the electric polarisation and the Kerr optical
function in a radio frequency a.c. field; while their electric susceptibility
function was given as a continued fraction, the Kerr function was in the form of
exponential integrals. Filippini \cite{filip72} measured experimentally the
Kerr dispersion constant when an a.c. field superimposed on a unidirectional
field is applied to a liquid. Coffey and Paranjape \cite{coffpara78}, Morita
\cite{morita78}, Morita and Watanabe \cite{mowa87}, gave theoretical
descriptions of these phenomena using pure classical diffusion equations. 

\subsection*{The electrical
susceptibility} In the classical limit, quantum equations reduce to the
classical HT equations for the electrical susceptibility \cite{tit96III}
  
\begin{equation}P(\tau)={\mu\over3}S_0^0(\tau), \label{eq:polar1}\end{equation}
 \begin{equation} 
\Bigl(\frac{d}{d\tau}+2j\Bigr)S_j^0(\tau)+2b_2\Bigl[(j+1)S_j^1(\tau)-jS_{j-1}^1(\tau)\Bigr]=0
\label{eq:sig0}\end{equation}  and \begin{equation} 
\Bigl(\frac{d}{d\tau}+2j+1\Bigr)S_j^1(\tau)-b_1\Bigl[S_j^0(\tau)-S_{j+1}^0
(\tau)\Bigr]=-b_1{\mu E_a\over k_BT}\Bigl(r+\cos\omega t\Bigr)\delta_{j,0},
\label{eq:sig01}\end{equation}where $\tau=Bt$, ${\omega}'=\omega/B$ are
dimensionless time and frequency, respectively; $r=E_c/E_a$ measures the ratio of
the constant field strength to the amplitude of the a.c. field.
$b_1b_2=\gamma=Ik_BT/\zeta^2,$ where $\zeta$ is the coupling coefficient. In the
steady state regime, we search for $S_j^m(\tau)$ in the forms: \begin{eqnarray} 
S_0^0(\omega',\tau)={\mu E_a\over k_BT}\Bigl[r+{S_0^0}'(\omega')e^{i{\omega}'
\tau}+({S_0^0}'(\omega'))^*e^{-i{\omega}'\tau}\Bigr],\nonumber\\ 
S_j^0({\omega}',\tau)={\mu E_a\over k_BT}{S_j^0}'(\omega')e^{i{\omega}'
\tau}+C.C. \mbox{  for $j\neq 0$},\nonumber\\S_j^1({\omega}',\tau)={\mu E_a\over
k_BT}{S_j^1}'(\omega')e^{i{\omega}' \tau}+C.C. \mbox{ for all
$j$}.\end{eqnarray} On substituting these into the hierarchy 
(\ref{eq:sig0}) -(\ref{eq:sig01}) and solving for ${S_0^0}'(\omega')$, we
get\begin{equation}
{S_0^0}'(\omega')=\displaystyle{\gamma\over{2\gamma+i\omega'\Biggl[
1+i\omega'+\displaystyle{2\gamma\over 2+i\omega'+\displaystyle{4\gamma\over
3+i\omega'+\displaystyle{4\gamma\over 4+i\omega'+\displaystyle{6\gamma\over
5+i\omega'+...}}}}\Biggr]}}\label{eq:s0}\end{equation}and using
(\ref{eq:polar1}), we deduce the polarisation \begin{equation}
P(\omega',\tau)={\mu^2 E_a\over 3k_BT} \Biggl\{{r\over 2}+\displaystyle{\gamma
e^{i\omega'\tau}\over{2\gamma+i\omega'\Biggl[
1+i\omega'+\displaystyle{2\gamma\over 2+i\omega'+\displaystyle{4\gamma\over
3+i\omega'+\displaystyle{4\gamma\over
4+i\omega'+...}}}\Biggr]}}+C.C.\Biggr\}.\end{equation} In the
absence of the d.c. field $(r=0)$, the result of Gross \cite{gross55} on
generalised Brownian motion is recovered. 
We define a reduced susceptibility $\chi_r(\omega',\tau)$
as \begin{equation}
\chi_r(\omega',\tau)=r+2\mid{S_0^0}'(\omega')\mid\cos\Bigl(\omega\tau-\alpha(\omega')\Bigr),\end{equation}
where $\alpha(\omega,)$, the phase difference between the exciting a.c. field
and the dielectric response function (the polarisation), furnishes valuable
informations on the absorption properties of the medium under investigation. It
is given by

\begin{equation} \tan\alpha(\omega')=-{{\it
Im}{S_0^0}'(\omega')\over{\it Re}{S_0^0}'(\omega')}.\end{equation} 
On neglecting inertial effects in (\ref{eq:s0}), we obtain the Debye
limit\begin{equation}{S_0^0}'(\omega')={1\over 2 (1+i\omega\tau_D)}\end
{equation}in usual frequency units. In this case, the phase is given by
$\tan\alpha(\omega)=\omega \tau_D$ with $\tau_D=\zeta/(2k_BT)$. The lowest
inertial limit, corresponding to the Rocard formula,
\begin{equation}{S_0^0}'(\omega')={1\over
2(1+i\omega\tau_D-(I\omega^2/2k_BT))}\end {equation}leads to the phase expression
 $\tan\alpha(\omega)=\omega \tau_D/\Bigl(1-(I\omega^2/2k_BT)\Bigr)$ which yields a maximum phase of $\pi/2$ for frequency of $\sqrt{2}$
 times mean thermal agitation frequency ($\omega_{mean}=\Bigl(k_BT/I\Bigr)^{0.5}$). At this frequency value, the rate of energy absorption from the surrounding bath by the rotators is in phase with the forcing field (since the rate of heat exchange between the rotator and the surrounding is proportional to minus the rate of change of the induced polarisation \cite{tit96II,tit96III,navthes95}).  On defining a new dimensionless frequency $\nu=\omega/\omega_{mean}$
in (\ref{eq:s0}), we rewrite ${S_0^0}'$
 as \begin{equation}{S_0^0}'(\nu)=\displaystyle{1\over
2+i\nu/\sqrt{\gamma}-\nu^2+\displaystyle{2i\sqrt{\gamma}\nu\over
2+i\sqrt{\gamma}\nu+ \displaystyle{4\gamma\over
3+i\sqrt{\gamma}\nu+\displaystyle{4\gamma\over 4+i\sqrt{\gamma}\nu+...}}}}.\end
{equation}

Figures 1 and 2 show the plots of the external exciting field $\cos
\nu t'$ and those of the  reduced susceptibility  $\chi_r(\omega',\tau)$ as functions of the dimensionless time $\tau$ (with $\tau=t\omega_{mean}$)
for $\nu=0.15$, $4.00$ and for different values of $\gamma$. For fixed $\omega_{mean}$, we analyse the effect of friction $\zeta$ on the phase,
through $\gamma=Ik_BT/\zeta^2$. Figure 3 shows a 3-D plot of $\chi_r(\omega',\tau)$ for $\gamma=0.05$.  Figure 4 shows a 3-D plot of the tangent of the phase angle as a function of the reduced frequency $\nu$ and $\gamma$. The resonance peaks are found to shift towards larger frequency values as $\gamma$ increases. It is important to note that while the Debye theory predicts such resonances only for infinite frequencies ($\tan\alpha(\omega)=\omega \tau_{D}$) and the Rocard lowest inertial limit predicts a resonance at $\nu=\sqrt{2}$, our results extensively portrays the effect of inertia on this resonance. 

\subsection*{The Kerr function}The classical HT equations \cite{tit96III} for the
optical Kerr function are:
  \begin{equation}\Phi(\tau)={1\over 30}Y_0^0(\tau)
\label{eq:kerrfunc}\end{equation}   
with
\begin{eqnarray} 
& &\Bigl(\frac{d}{d\tau}+2j\Bigr)Y_j^0(\tau)+24b_2\Bigl((j+1)Y_j^1(\tau)-
jY_{j-1}^1(\tau)\Bigr)\nonumber\\& &=-4b_2{\mu E_a\over k_BT}
(r+\cos\omega'\tau)S_{j-1}^1(\tau)(1-\delta_{j,0}),\label{eq:ker2}\end{eqnarray}

\begin{eqnarray}  & &\Bigl(\frac{d}{d\tau}+2j+1\Bigr)Y_j^1(\tau)-{b_1\over
3}\Bigl(Y_j^0(\tau)- Y_{j+1}^0(\tau)\Bigr)+{b_1\over
3}\Bigl(X_j(\tau)- X_{j+1}(\tau)\Bigr)\nonumber\\ & &=-b_1{\mu E_a\over
k_BT}(r+\cos\omega'\tau)S_j^0(\tau)-b_1{\Delta\alpha E_a^2\over
k_BT}(r+\cos\omega'\tau)^2 \delta_{j,0},\label{eq:ker3}\end{eqnarray}
\begin{eqnarray}  &
&\Bigl[(2j+1)\Bigl(\frac{d}{d\tau}+2j\Bigr)+2\Bigr]X_j(\tau)-j\Bigl(\frac{d}
{d\tau}+2j-2\Bigr)X_{j-1}(\tau)\nonumber\\ &
&-(j+1)\Bigl(\frac{d}{d\tau}+2j+2\Bigr)X_{j+1} (\tau)-{1\over 2}
Y_j^0(\tau)=b_2{\mu E_a\over k_BT}(r+\cos\omega' \tau)\nonumber\\ & &\times
\Bigl[-2j(j-1)S_{j-2}^1(\tau)+j(4j+5)S_
{j-1}^1(\tau)-(j+1)(2j+3)S_j^1(\tau)\Bigr]. \label{eq:ker1}\end{eqnarray}  
Steady state solutions are sought in the forms:
\begin{eqnarray}
& &X_j(\omega',\tau)=\Bigl({\mu E_a\over
k_BT}\Bigr)^2\Bigl[X_{j,0}(\omega')+X_{j,1}(\omega')e^{i\omega'\tau}+X_{j,2}(\omega')
e^{2i\omega'\tau} +C.C.\Bigr],\nonumber\\
& &Y_j^0(\omega',\tau)=\Bigl({\mu E_a\over
k_BT}\Bigr)^2\Bigl[Y_{j,0}^0(\omega')+Y_{j,1}^0(\omega')e^{i\omega'\tau}+Y_{j,2}^0
(\omega')e^{2i\omega'\tau} +C.C. \Bigr],\nonumber\\
& &Y_j^1(\omega',\tau)=\Bigl({\mu E_a\over
k_BT}\Bigr)^2\Bigl[Y_{j,0}^1(\omega')+Y_{j,1}^1(\omega')e^{i\omega'\tau}+Y_{j,2}^1
(\omega')e^{2i\omega'\tau} +C.C.\Bigr], \end{eqnarray} where C.C. denotes
complex conjugate. Knowing the forms of
$S_j^m$, we obtain the three systems of hierarchies (each system being a set of
three coupled equations (triplets)):

t1) \begin{eqnarray} &
&2jY_{j,0}^0(\omega')+24b_2\Bigl((j+1)Y_{j,0}^1(\omega')-
jY_{j-1,0}^1(\omega')\Bigr)=-2b_2S_{j-1}^1(\omega')(1-\delta_{j,0}), 
\newline\nonumber\\& &(2j+1)Y_{j,0}^1(\omega')-{b_1\over
3}\Bigl(Y_{j,0}^0(\omega')- Y_{j+1,0}^0(\omega')\Bigr)+{b_1\over
3}\Bigl(X_{j,0}(\omega')- X_{j+1,0}(\omega')\Bigr)\nonumber\\&
&=-{b_1\over 4}\Bigl[(2r^2+2r^2/R+1/R)\delta_{j,0}+2S_j^0(\omega') \Bigr],
\newline\nonumber\\&
&\Bigl[2j(2j+1)+2\Bigr]X_{j,0}(\omega')-j(2j-2)X_{j-1,0}(\omega)
-(j+1)(2j+2\Bigr)X_{j+1,0} (\omega')\nonumber\\& &-1/2 Y_{j,0}^0(\omega') 
={b_2\over 2}\Bigl[-2j(j-1)S_{j-2}^1(\omega')+j(4j+5)S_
{j-1}^1(\omega')\nonumber\\& &-(j+1)(2j+3)S_j^1(\omega')\Bigr]; \end{eqnarray}

t2) \begin{eqnarray} &
&\Bigl(i\omega'+2j\Bigr)Y_{j,1}^0(\omega')+24b_2\Bigl((j+1)
Y_{j,1}^1(\omega')-jY_{j-1,1}^1(\omega')\Bigr)\nonumber\\& &=
-4rb_2S_{j-1}^1(\omega')(1- \delta_{j,0}),\newline\nonumber\\&
&\Bigl(i\omega'+2j+1\Bigr)Y_{j,1}^1(\omega') -{b_1\over
3}\Bigl(Y_{j,1}^0(\omega')- Y_{j+1,1}^0(\omega')\Bigr)\nonumber\\& &+{b_1\over
3}\Bigl(X_{j,1}(\omega')-
X_{j+1,1}(\omega')\Bigr)=-r{b_1\over 4}\Bigl[(2+4/R)\delta_{j,0}+4S_j^0\Bigr],
\newline\nonumber\\&
&\Bigl[(2j+1)\Bigl(i\omega'+2j\Bigr)+2\Bigr]X_{j,1}(\omega')
-j\Bigl(i\omega'+ 2j-2\Bigr)X_{j-1}(\omega')\nonumber\\& &
-(j+1)\Bigl(i\omega'+2j+2\Bigr)X_{j+1,1} (\omega')-(1/2)
Y_{j,1}^0(\omega')=b_2r \Bigl[-2j(j-1)S_{j-2}^1(\omega')\nonumber\\& &+j(4j+5)S_
{j-1}^1(\omega')-(j+1)(2j+3)S_j^1(\omega')\Bigr]  \end{eqnarray} and

t3) \begin{eqnarray}& &\Bigl(2i\omega'+2j\Bigr)Y_{j,2}^0(\omega')+
24b_2\Bigl((j+1)Y_{j,2}^1
(\omega')-jY_{j-1,2}^1(\omega')\Bigr)\nonumber\\& &
=-2b_2S_{j-1}^1(\omega')(1-\delta_{j,0}), 
\newline\nonumber\\& &\Bigl(2i\omega'+2j+1\Bigr)Y_{j,2}^1(\omega')-{b_1\over
3}\Bigl(Y_{j,2}^0(\omega')- Y_{j+1,2}^0(\omega')\Bigr)\nonumber\\& &+{b_1\over
3}\Bigl(X_{j,2}(\omega')-
X_{j+1,2}(\omega')\Bigr)=-{b_1\over 4}\Bigl[{1\over
R}\delta_{j,0}+2S_j^0\Bigr],  \newline\nonumber\\& &
\Bigl[(2j+1)\Bigl(2i\omega'+2j\Bigr)+2\Bigr]X_{j,2}(\omega')-j\Bigl(2i\omega'+
2j-2\Bigr)X_{j-1,2}(\omega')\nonumber\\& &-(j+1)\Bigl(2i\omega'+2j+
2\Bigr)X_{j+1,2}   (\omega')-{1\over 2}Y_{j,2}^0(\omega') ={b_2\over 2}
\Bigl[-2j(j-1)S_{j-2}^1(\omega')\nonumber\\& &+j(4j+5)S_
{j-1}^1(\omega')-(j+1)(2j+3)S_j^1(\omega')\Bigr],\end{eqnarray}where
$R=\mu^2/(\Delta\alpha k_BT)$. For simplicity, we have left out the primes on
each $S_j^m$. $i$ denotes the complex number $\sqrt{-1}$. 

The technique adopted in solving these triplets is based on convergents. Remark
that the above systems could be written in matrix forms of infinite
dimensions. The notion of convergence can be seen as limiting the dimensions of
the matrices. The zeroth convergent consists of considering only equations
involving just $j=0$. The first convergent is the modification of the zeroth
by including $j=1$ terms. The former is the solution of a $3\times
3$ matrix equation, while the latter is that of a $6\times 6$ matrix equation.
Reliable spectral informations can only be got from at least a $6\times 6$ matrix
equation. The following are the expressions for $Y_{0,0}^0(\omega')$,
$Y_{0,1}^0(\omega')$ and $Y_{0,2}^0(\omega')$ obtained for $j=1$:

\begin{eqnarray}Y_{0,0}^0(\omega')=& &{1 \over 2}(\alpha+2r-\alpha r)+(1-\alpha)(1-r)\Biggl\{(1-{i\omega' \over 2})\nonumber\\& &+{2i\omega'\over 3(2+5\gamma)}\Biggl[2-\gamma-\displaystyle{8\gamma\over
2+i\omega'+\displaystyle{4\gamma\over 3+i\omega'+\displaystyle
{4\gamma\over 4+i\omega'+...}}}
\nonumber\\& &+\displaystyle{4\gamma^2/3\over
4\gamma+(2+i\omega')\Bigl(3+i\omega'+\displaystyle{4\gamma\over 4+i\omega'+
\displaystyle{6\gamma\over
5+i\omega'+\displaystyle{6\gamma\over
6+i\omega'+...}}}\Bigr)}\Biggr]\Biggr\}
\nonumber\\& &
\times \displaystyle{\gamma\over
2\gamma+i\omega'\Bigl(1+i\omega'+\displaystyle{2\gamma\over 2+i\omega'+
\displaystyle{4\gamma\over 3+i\omega'+\displaystyle{4\gamma\over
4+i\omega'+...}}}\Bigr)}\nonumber\\& &+{64\over 9}
{\gamma\over 2+5\gamma}Y_{2,0}^0(\omega'),
\end{eqnarray}
 
\begin{eqnarray}Y_{0,1}^0(\omega')=
& &\sqrt{r(1-r)}\Biggl\{(1+2/R){6\gamma (1+\alpha)\over
1+i\omega'}\Bigl(2+i\omega'+4\gamma{15+4i\omega'\over
(3+i\omega')(4+i\omega')}\Bigr)\nonumber\\&+&
(1-\alpha)\Biggl[{24\gamma\over(1+i\omega')(2+i\omega')}\Bigl(2+i\omega'+
4\gamma{15+4i\omega'\over(3+i\omega')(4+i\omega')}\Bigr)\nonumber\\&+&
{8i\omega'\gamma\over 1+i\omega'}-{32i\omega'\gamma^2\over
(1+i\omega')(3+i\omega')(4+i\omega')}\Biggl({6(1+i\omega')\over(2+i\omega')}
\nonumber\\&-&\displaystyle{4\gamma\over
4\gamma+(2+i\omega')\Bigl(3+i\omega'+\displaystyle{4\gamma\over 4+i\omega'+
\displaystyle{6\gamma\over 5+i\omega'+\displaystyle{6\gamma\over
6+i\omega'+...}}}\Bigr)}\Biggr)-\nonumber\\&-& 
\displaystyle{192\gamma^2i\omega'\over
(1+i\omega')(3+i\omega')\Bigl(2+i\omega'+\displaystyle{4\gamma\over3+i\omega'+
\displaystyle{4\gamma\over 4+i\omega'+\displaystyle{6\gamma\over
5+i\omega'+...}}}\Bigr)}\Biggr]\nonumber\\&\times&\displaystyle{\gamma\over
2\gamma+i\omega'\Bigl(1+i\omega'+\displaystyle{2\gamma\over2+i\omega'+
\displaystyle{4\gamma\over 3+i\omega'+\displaystyle{4\gamma\over
4+i\omega'+...}}}\Bigr)}\nonumber\\&+&{128\gamma^2/r\over
3+i\omega'}Y_{2,1}^0(\omega')\Biggr\}\bigg/
\Biggl\{\Bigl(i\omega'+4\gamma{3+i\omega'\over
(1+i\omega')(2+i\omega')}\Bigr)
\Bigl(2+i\omega'\nonumber\\&+&4\gamma{15+4i\omega'\over
(3+i\omega')(4+i\omega')}\Bigr)+{8i\omega'\gamma\over
1+i\omega'}\nonumber\\&-&{32i\omega'\gamma^2\over
(1+i\omega')(2+i\omega')(3+i\omega')(4+i\omega')} \Biggr\},\end{eqnarray}

\begin{eqnarray}Y_{0,2}^0(\omega')=& &(1-r)\Biggl\{{3\gamma\alpha \over
1+2i\omega'}\Bigl(2+2i\omega'+4\gamma{15+8i\omega'\over
(3+2i\omega')(4+2i\omega')}\Bigr)\nonumber\\&+&
(1-\alpha)\Biggl[6\gamma{2+i\omega'\over(1+2i\omega')(2+2i\omega')}\Bigl(2+2i\omega'+
4\gamma{15+8i\omega'\over(3+2i\omega')(4+2i\omega')}\Bigr)\nonumber\\&+&{8i\omega'\gamma\over
1+2i\omega'}-{96i\omega'\gamma^2\over
(1+2i\omega')(3+2i\omega')\Bigl(2+i\omega'+\displaystyle{4\gamma\over
3+i\omega'+ \displaystyle{4\gamma\over
4+i\omega'+...}}\Bigr)}\nonumber\\&-&16i\omega'\gamma^2{\Biggl(3-\displaystyle
{4\gamma\over
4\gamma+(2+i\omega')\Bigl(3+i\omega'+\displaystyle{4\gamma\over 4+i\omega'+
...}\Bigr)}+6i\omega'\Biggr)
\over(1+2i\omega')(2+2i\omega')(3+2i\omega')(4+2i\omega') }\Biggr]
\nonumber\\& \times& \displaystyle{\gamma\over
2\gamma+i\omega'\Bigl(1+i\omega'+\displaystyle{2\gamma\over2+i\omega'+
\displaystyle{4\gamma\over 3+i\omega'+\displaystyle{4\gamma\over
4+i\omega'+...}}}\Bigr)}+\nonumber\\&+&
{128\gamma^2\over(1+2i\omega')(3+2i\omega')} Y_{2,2}^0(\omega')\Biggr\}
\bigg/\Biggl\{\Bigl[2i\omega'+4\gamma{3+4i\omega'\over
(1+2i\omega')(2+2i\omega')}\Bigr]\times\nonumber\\&\times&\Bigl[2+2i\omega'+4\gamma{15+8i\omega'\over
(3+2i\omega')(4+2i\omega')}\Bigr]\nonumber\\&+&{16i\omega'\gamma\Bigl((2+i\omega')
(3+i\omega')(4+i\omega')-4\gamma\Bigr)\over
(1+2i\omega')(2+2i\omega')(3+2i\omega')(4+2i\omega')} \Biggr\}.\label{eq:kercl} \end {eqnarray}
We can now write the Kerr function as
\begin{equation}\Phi(\omega',\tau)={1\over 30}E^2 K_0
\Bigl[Y_{0,0}^0(\omega')+Y_{0,1}^0(\omega')\exp(i\omega' \tau)+
Y_{0,2}^0(\omega')\exp(2i\omega' \tau)+C.C.\Bigr],\end{equation}
 where $E^2=E_a^2+E_c^2$ and $K_0=\Bigl({\mu \over K_BT}\Bigr)^2+{\Delta \alpha \over K_BT}$. Remark that in the last expressions we have replaced the field parameter $r=E_c/E_a$ with a more convenient one $r=E_c^2/E^2$ and the quantity $R$ is replaced by $\alpha= {\Delta \alpha/(k_BT) \over K_0}$. With these new parameters, the limiting cases are better understood, for example $r=0$ corresponds to pure a.c. field effects and $\alpha=0$ demonstrates the properties of a non polarisable but polar molecule. Note that both parameters are such that $0\le r\le 1$ and $0 \le \alpha \le 1$. The Kerr function (eq.\ref{eq:kercl}) presents very interesting properties. It expresses the time, radio frequency (rf), rotator-bath parameters and more importantly the $E_c/E_a$
dependences of the Kerr electrical birefringence (KEB). For infinitely high
frequencies, the function reduces to the constant field steady state expression

\begin{equation}\Phi_\infty={E_c^2\over15}\Bigl(({\mu^2\over
k_BT})^2+{\Delta\alpha\over k_BT}\Bigr)+{E_a^2\over30}{\Delta\alpha\over
k_BT}.\end{equation}The effect of the a.c. field is felt only when
$\Delta\alpha=\alpha_\parallel-\alpha_\bot\neq 0$. This result is consistent
with that of Doi and Edwards \cite{doi}. At very high frequencies,
a.c. field effects on dipole moments average
out. Also, in the absence of the d.c. field ($r=0$), the term
$Y_{0,1}^0(\omega')$ vanishes and the result for pure a.c. field is recovered.

\subsection{The rotating wave approximation (RWA) limit }  In this limit, the
solution of the rotators in the bath is assumed highly diluted, the
pressure and friction are very low. The coupling parameter $B$  or
the characteristic rotator-bath frequency is very small compared to the rotator
lines $\omega_l=(\hbar l/I)$. The dynamics of the rotator is mainly governed
by free  rotations and  interactiions with  the re-orienting fields. Bath coupling affects
only the frequency shifts and line widths. The absorption lines are the neat
spectral lines corresponding to the different $l$ transitions owing to non
negligible Planck constant $\hbar$ and finite inertia \cite{gross55}. The
transition frequencies are  
$$\omega_{l->l\pm \Delta l}=\mid(E_{l\pm 
\Delta l}-E_l)\mid /\hbar=(2l+1+\Delta l)\hbar/2I $$ for transitions from $l$ to
$l\pm\Delta l$. At the level of linear response, $ \Delta l=1$ and
$\omega_{l+1}=(l+1)\hbar/I$. For the lowest order nonlinear effect (the Kerr
effect to the second order in electric field), $ \Delta l=2$ and
$\omega_{2l+3}=(2l+3)\hbar/I$.

The relevant dielectric matrix elements are governed by the quantum HT
equations for the electrical susceptibility and the Kerr optical functions
\cite{tit96III}:

   \begin{equation} \Bigl({\partial\over\partial
t}-i(\omega_{l+1}+ \Delta\omega_{l+1})+\Gamma_{l+1}\Bigr)\sigma_
{l,l+1}(t)=-i{\mu E_a\over \hbar}(r+\cos\omega t)\Bigl(1-e^{-\beta
\bigl(E_{l+1}-E_l\bigr)}\Bigr), \label{eq:mast1} \end{equation} 
 
\begin {equation}
\Bigl(\frac{\partial}{\partial t}+\gamma_l\Bigr)\varphi_{l,l}(t)={\mu E_a
\over \hbar}(r+\cos\omega t)\Biggl({2l-1 \over
2l+1}Im\sigma_{l,l+1}(t)-e^{\beta\bigl(E_l-E_{l-1}\bigr)}{2l+3\over
2l+1}Im\sigma_{l-1,l}(t)(1-\delta_{l,0})\Biggr),\label{eq:mast2} \end{equation}

\begin {eqnarray} & & \Bigl({\partial\over\partial t}-i(\omega_{2l+3}+
\Delta\omega_{2l+3})+\Gamma_{2l+3}\Bigr)\eta_ {l,l+2}(t)=i{\mu E_a
\over \hbar}(r+\cos\omega t)\Biggl(e^{-\beta \bigl( E_{l+1}-E_l\bigr)}\sigma_{l+1,l+2}(t)\nonumber\\&
&-\sigma_{l,l+1}(t)\Biggr) -i{\Delta \alpha E_a^2\over
2\hbar}(r+\cos\omega t)^2\Biggl(1-e^{-\beta\bigl(E_{l+2}-E_l\bigr)}
\Biggr).\label{eq:mast3} \end{eqnarray}
 Here, we define the dimensionless line widths and
frequency shifts:  
\begin {equation} \gamma_l'={I\over
\hbar}\gamma_l={2BI\over\hbar}{1 \over 2l+1} \Biggl [l^2\Bigl(1+{1\over
e^{\beta\hbar\omega_l}-1}\Bigr)+ {(l+1)^2\over
e^{\beta\hbar\omega_{l+1}}-1}\Biggr], \label{eq:width}\end{equation} 

\begin{equation} \Gamma_{l+1}'={1\over 2}\Bigl(\gamma_l'+\gamma_{l+1}'\Bigr),
\end{equation} \begin {equation} \Gamma_{2l+3}'={1\over
2}\Bigl(\gamma_l'+\gamma_{l+2}'\Bigr), \end{equation}  

\begin {eqnarray}
\Delta\omega_{l+1}'&=&-{4\hbar^3B\over
Ik_B^2T^2}(2l+3)\sum_{n=0}^\infty{(2n\pi)^3 \over
\Bigl[(2n\pi)^2+(al)^2\Bigr]}\nonumber\\& &\times {1\over\Bigl[(2n\pi)^2+
(a(l+1))^2\Bigr]\Bigl[(2n\pi)^2+(a(l+2))^2\Bigr]}\label{eq:shift}\end{eqnarray} 
and
\begin {eqnarray} \Delta\omega_{2l+3}'&=&-{4\hbar^3B\over
Ik_B^2T^2}(2l+3)\sum_{n=0}^\infty
{(2n\pi)^5\Bigl(1+{a\over(n\pi)^2}(l^2+3l+3)\Bigr) \over
\Bigl[(2n\pi)^2+(al)^2\Bigr]\Bigl[(2n\pi)^2+(a(l+1))^2 \Bigr]}\nonumber\\&
&\times {1\over\Bigl[(2n\pi)^2+(a(l+2))^2\Bigr]
\Bigl[(2n\pi)^2+(a(l+3))^2\Bigr]},\end{eqnarray} where $a=\hbar^2/Ik_BT$.
These functions well indicate how line widths  and frequency shifts respond to
changing physical parameters like inertia, friction and temperature, thus their
utility in exploring the influence of the  parameter variations on spectral
lines. Note that, in our dimensionless frequency units,  we define the quantum
state frequency $\omega_l'=l$. 

 \subsection*{The electrical susceptibility} We are interested in the steady
state regime. On solving equation (\ref{eq:mast1}) for this, we get 
\begin{eqnarray}
& &\sigma_{l,l+1}^{st}(\omega, t)=
(\mu E_a/\hbar)\Bigl(1-\exp[-{\hbar^2\over 
Ik_BT}(l+1)]\Bigr)\Biggl\{{r\over\omega_{l+1}+\Delta\omega_{l+1}+i\Gamma_{l+1}}
\nonumber\\& &+{e^{i\omega t}\over 2(\omega_{l+1}+\Delta\omega_{l+1}-\omega+i\Gamma_{l+1})}
+{e^{-i\omega t}\over
2(\omega_{l+1}+\Delta\omega_{l+1}+\omega+i\Gamma_{l+1})}\Biggr\}.\label{eq:qsig}\end{eqnarray}  
The polarisation is deduced as  
\begin{eqnarray}
P(\omega, t)&=&{\mu^2 E_a\over
3k_BT}\sum_{l=0}^\infty\Bigl(e^{-\beta E_l}-e^{-\beta E_{l+1}} \Bigr){l+1 \over
a}\Bigl(l+1+\Delta\omega_{l+1}'\Bigr)\nonumber\\ &
&\times\Biggl\{r\bigg/\Bigl[\Bigl(l+1+\Delta\omega_{l+1}'\Bigr)^2+
\Gamma_{l+1}'^2\Bigr]+\Biggl(\Bigl[\Bigl(l+1+\Delta\omega_{l+1}'\Bigr)^2-\omega'^2+
\Gamma_{l+1}'^2\Bigr]\cos\omega t\nonumber\\& &+2\Gamma_{l+1}'\omega'\sin\omega
t\Biggr)\bigg/\Biggl(\Bigl[\Bigl(l+1+\Delta\omega_{l+1}'\Bigr)^2-\omega'^2+
\Gamma_{l+1}'^2\Bigr]^2+4\omega'^2\Gamma_{l+1}'^2\Biggr)\Biggr\}.\end{eqnarray}
These are the Van Vleck-Weisskopf line forms for the electrical susceptibility.
Sharp separate lines result for small widths at half heights $\Gamma_{l+1}$. For
line coupling and subsequent line overlaps to be absent, thus, $\Gamma_{l+1}$
should be small compared to line spacings which for the electrical
susceptibility stands at $\hbar/I$. Remark that the Boltzmann weight $e^{-\beta
E_l}$ appearing in the last expression renders small quantum number
transitions more probable. An appropriate Taylor expansion of the Bose-Einstein
factor appearing in the expression of the half width shows that a necessary
condition for dominant lines is expressed by the inequality
$(B/\omega_{mean})^2<<a^3/4=(\hbar^2/Ik_BT)^3/4$.

For $r=0$, we define the
reduced susceptibility     \begin{equation} \chi_r(\omega,t)=\cos\Bigl(\omega
t-\alpha(\omega)\Bigl) \end{equation}
where $\alpha(\omega)$, the phase difference between the exciting field and
the induced polarisation, is given by
\begin{eqnarray}\tan\alpha(\omega)&=&\Biggl[\sum_{l=0}^\infty\Bigl(e^{-\beta
E_l}-e^{-\beta
E_{l+1}}\Bigr)(l+1)\Bigl(l+1+\Delta\omega_{l+1}'\Bigr)\Gamma_{l+1}'
\omega'\bigg/\nonumber\\& &
\Biggl(\Bigl[\Bigl(l+1+\Delta\omega_{l+1}'\Bigr)^2-\omega'^2+
\Gamma_{l+1}'^2\Bigr]^2+4\omega'^2\Gamma_{l+1}'^2\Biggr)\Biggr]\bigg/\nonumber\\&
&\Biggl[\sum_{l=0}^\infty\Bigl(e^{-\beta E_l}-e^{-\beta E_{l+1}}
\Bigr)(l+1)\Bigl(l+1+\Delta\omega_{l+1}'\Bigr)\Bigl(\Bigl(l+1+\Delta\omega_{l+1}'\Bigr)^2\nonumber\\& &-\omega'^2+
\Gamma_{l+1}'^2\Bigr)\bigg/\Biggl(\Bigl[\Bigl(l+1+\Delta\omega_{l+1}'\Bigr)^2-\omega'^2+
\Gamma_{l+1}'^2\Bigr]^2+4\omega'^2\Gamma_{l+1}'^2\Biggr)\Biggr].\end{eqnarray}
Note that for usual temperatures and simple linear molecules like HCl and DCl
\cite{mowaca76,starck}, the frequency shift has a negligible contribution as it
varies as $\Delta\omega_{l+1}'\sim-10^{-4}({2l+3})$ compared to the corresponding
line $l+1$.  

\subsection*{The Kerr function}On using the
 expression for $\sigma_{l,l+1}^{st}(\omega,t)$ (equation
(\ref{eq:qsig})) into equations (\ref{eq:mast2}) and (\ref{eq:mast3}) and
solving the resulting equations for the steady state matrix elements
$\varphi_{l,l}^{st}(\omega,t)$ and $\eta_{l,l+2}^{st}(\omega,t)$, we deduce
that the Kerr function comprises three terms: a frequency dependent time
constant term $\Phi_0(\omega)$, an $\omega-$frequency time dependent term 
$\Phi_1(\omega)e^{i\omega t}$ and a $2\omega-$ frequency time dependent one
$\Phi_2(\omega)e^{2i\omega t}$ with their respective complex conjugates. In
other words,
\begin{equation}
\Phi(\omega,t)=\Phi_0(\omega)+\Phi_1(\omega)e^{i\omega
t}+\Phi_2(\omega)e^{2i\omega t} +C.C.\label{eq:keb},
\end{equation}
where
 \begin{eqnarray}
\Phi_0(\omega')&=&{1\over 60}E^2 K_0\sum_{l=0}^\infty{l+1\over a }\Biggl[(1-\alpha){l\over
2l-1}{1\over\gamma_l'}\Bigl(e^{-\beta E_{l-1}}-e^{-\beta E_l}\Bigr)\nonumber\\&
&\times\Biggl(2r{\Gamma_l'+i(l+\Delta\omega_l')\over
(l+\Delta\omega_l')^2+\Gamma_l'^2}+{1\over
2}(1-r){\Gamma_l'+i(l+\Delta\omega_l'-\omega')\over
(l+\Delta\omega_l'-\omega')^2+\Gamma_l'^2}\Biggl)\nonumber\\& &-{l\over
2l+3}(1-\alpha){1\over\gamma_l'}\Bigl(e^{-\beta E_l}-e^{-\beta
E_{l+1}}\Bigr)\Biggl(2r{\Gamma_{l+1}'+i(l+1+\Delta\omega_{l+1}')\over
(l+1+\Delta\omega_{l+1}')^2+\Gamma_{l+1}'^2}\nonumber\\& &+{1\over
2}(1-r){\Gamma_{l+1}'+i(l+1+ \Delta\omega_{l+1}'-\omega')\over
({l+1}+\Delta\omega_{l+1}'-\omega')^2+\Gamma_{l+1}'^2}\Biggl)+{3(l+2)\over 2l+3}{1\over (2l+3+\Delta\omega_{2l+3}')^2+\Gamma_{2l+3}'^2}\nonumber\\&
&\times\Biggl\{\Bigl(2l+3+\Delta\omega_{2l+3}')-i\Gamma_{2l+3}'\Bigr)
\Biggl[\Bigl(e^{-\beta E_l}-e^{-\beta E_{l+1}}\Bigr)\nonumber\\& &\times 
\Biggl(2r{(l+1+\Delta\omega_{l+1}'-i\Gamma_{l+1}')\over
(l+1+\Delta\omega_{l+1}')^2+\Gamma_{l+1}'^2}+{1\over
2}(1-r){(l+1+ \Delta\omega_{l+1}'-\omega'-i\Gamma_{l+1}')\over
(l+1+\Delta\omega_{l+1}'-\omega')^2+\Gamma_{l+1}'^2}\nonumber\\& &
+{1\over 2}(1-r){(l+1+ \Delta\omega_{l+1}'+\omega'-i\Gamma_{l+1}')\over
({l+1}+\Delta\omega_{l+1}'+\omega')^2+\Gamma_{l+1}'^2}\Biggl)    - 
\Bigl(e^{-\beta E_{l+1}}-e^{-\beta E_{l+2}}\Bigr)\nonumber\\& &\times 
\Biggl(2r{(l+2+\Delta\omega_{l+2}'-i\Gamma_{l+2}')\over
(l+2+\Delta\omega_{l+2}')^2+\Gamma_{l+2}'^2}+{1\over
2}(1-r){(l+2+ \Delta\omega_{l+2}'-\omega'-i\Gamma_{l+2}')\over
(l+2+\Delta\omega_{l+2}'-\omega')^2+\Gamma_{l+2}'^2}\nonumber\\& &
+{1\over 2}(1-r){(l+2+ \Delta\omega_{l+2}'+\omega'-i\Gamma_{l+2}')\over
({l+2}l+\Delta\omega_{l+2}'+\omega')^2+\Gamma_{l+2}'^2}\Biggl)\Biggr](1-\alpha)\nonumber\\&
&+{\alpha\over 2}\Bigl(e^{-\beta E_l}-e^{-\beta
E_{l+2}}\Bigr) (2l+3+ \Delta\omega_{2l+3}')\Biggr\}\Biggr],  \end{eqnarray}

\begin{eqnarray}
\Phi_1(\omega')&=&{\sqrt{r(1-r)}\over 60}E^2K_0\sum_{l=0}^\infty{l+1\over a
}\Biggl[{2(1-\alpha)(\gamma_l'-i\omega')\over\gamma_l'^2+\omega'^2}\Biggl\{{l\over
2l+3}\Bigl(e^{-\beta E_l}-e^{-\beta
E_{l+1}}\Bigr)\nonumber\\& &\times\Biggl({\Gamma_{l+1}'\over
(l+1+\Delta\omega_{l+1}')^2+\Gamma_{l+1}'^2}+{\Gamma_{l+1}'+i(l+1+
\Delta\omega_{l+1}'+\omega')\over
(l+1+\Delta\omega_{l+1}'+\omega')^2+\Gamma_{l+1}'^2}-\nonumber\\& &-
{\Gamma_{l+1}'+i(l+1+ \Delta\omega_{l+1}'-\omega')\over
(l+1+\Delta\omega_{l+1}'-\omega')^2+\Gamma_{l+1}'^2}\Biggl)-{l\over
2l-1}\Bigl(e^{-\beta E_{l-1}}-e^{-\beta
E_l}\Bigr)\nonumber\\& &\times\Biggl({\Gamma_l'\over
(l+\Delta\omega_l')^2+\Gamma_l'^2}+{\Gamma_l'+i(l+
\Delta\omega_l'+\omega')\over
(l+\Delta\omega_l'+\omega')^2+\Gamma_l'^2}\nonumber\\& &- {\Gamma_l'+i(l+
\Delta\omega_l'-\omega')\over
(l+\Delta\omega_l'-\omega')^2+\Gamma_l'^2}\Biggl)\Biggr\}+ {3(l+2)\over
2l+3}\nonumber\\& &
\times{\Bigl(2l+3+\Delta\omega_{2l+3}'\Bigr)^2-\omega'^2-2i\omega'\Gamma_{2l+3}'\over\Bigl[(2l+3+\Delta\omega_{2l+3}')^2-\omega'^2+\Gamma_{2l+3}'^2\Bigr]^2
+4\omega'^2\Gamma_{2l+3}'^2}\Biggl\{(1-\alpha)\Bigl(2l+3+\Delta\omega_{2l+3}'
\nonumber\\& &+\omega'-i\Gamma_{2l+3}'\Bigr)
\Biggl[\Bigl(e^{-\beta E_l}-e^{-\beta
E_{l+1}}\Bigr)\Biggl({l+1+\Delta\omega_{l+1}'-\omega'-i\Gamma_{l+1}'\over
(l+1+\Delta\omega_{l+1}'-\omega')^2+\Gamma_{l+1}'^2}\nonumber\\&
&+{l+1+\Delta\omega_{l+1}'-i\Gamma_{l+1}'- \over
(l+1+\Delta\omega_{l+1}')^2+\Gamma_{l+1}'^2}\Biggr)-\Bigl(e^{-\beta
E_{l+1}}-e^{-\beta E_{l+2}}\Bigr)\nonumber\\& &\times\Biggl({(l+2+\Delta
\omega_{l+2}'-\omega'-i\Gamma_{l+2}')\over
(l+2+\Delta\omega_{l+2}'-\omega')^2+\Gamma_{l+2}'^2}+{l+2+\Delta\omega_{l+2}'
-i\Gamma_{l+2}'\over (l+2+\Delta\omega_{l+2}')^2+\Gamma_{l+2}'^2}
\Biggr)\Biggr]\nonumber\\& &+\Bigl(2l+3+\Delta\omega_{2l+3}'
-\omega'+i\Gamma_{2l+3}'\Bigr) \Biggl[\Bigl(e^{-\beta E_{l}}-e^{-\beta
E_{l+1}}\Bigr)\nonumber\\&
&\times\Biggl({l+1+\Delta\omega_{l+1}'+\omega'+i\Gamma_{l+1}'\over
(l+1+\Delta\omega_{l+1}'-\omega')^2+\Gamma_{l+1}'^2}+
{l+1+\Delta\omega_{l+1}'+i\Gamma_{l+1}'\over(l+1+\Delta\omega_{l+1}')^2+
\Gamma_{l+1}'^2}\Biggr)\nonumber\\& &-\Bigl(e^{-\beta E_{l+1}}-e^{-\beta
E_{l+2}}\Bigr)\Biggl({l+2+\Delta \omega_{l+2}'+\omega'+i\Gamma_{l+2}'\over
(l+2+\Delta\omega_{l+2}'-\omega')^2+\Gamma_{l+2}'^2}\nonumber\\&
&+{l+2+\Delta\omega_{l+2}'
+i\Gamma_{l+2}'\over(l+2+\Delta\omega_{l+2}')^2+\Gamma_{l+2}'^2}
\Biggr)\Biggr](1-\alpha)\nonumber\\& &+\alpha (2l+3+\Delta\omega_{2l+3}')
\Bigl(e^{-\beta E_l}-e^{-\beta E_{l+2}}\Bigr)\Biggr\}\Biggr],     \end{eqnarray}

\begin{eqnarray}
\Phi_2(\omega')&=&{(1-r)\over 60}E^2K_0\sum_{l=0}^\infty{l+1\over a
}\Biggl[{2(1-\alpha)(\gamma_l'-2i\omega')\over\gamma_l'^2+4\omega'^2}\Biggl\{{l\over
2l+3}\Bigl(e^{-\beta E_l}-e^{-\beta
E_{l+1}}\Bigr)\nonumber\\&
&\times\Biggl({\Gamma_{l+1}'+i(l+1+\Delta\omega_{l+1}'+\omega')\over
(l+1+\Delta\omega_{l+1}'+\omega')^2+\Gamma_{l+1}'^2}-{\Gamma_{l+1}'+i(l+1+
\Delta\omega_{l+1}'-\omega')\over
(l+1+\Delta\omega_{l+1}'-\omega')^2+\Gamma_{l+1}'^2}\Biggl)\nonumber\\&
&-{l\over 2l-1}\Bigl(e^{-\beta E_{l-1}}-e^{-\beta
E_l}\Bigr)\Biggl[{(\Gamma_l'+i(l+\Delta\omega_l'+\omega')\over
(l+\Delta\omega_l'+\omega')^2+\Gamma_l'^2}-\nonumber\\& &-{\Gamma_l'+i(l+
\Delta\omega_l'-\omega')\over (l+\Delta\omega_l'-\omega')^2+\Gamma_l'^2}
\Biggl]\Biggr\}+ {3(l+2)\over 2l+3}\nonumber\\& &
\times({2l+3+\Delta\omega_{2l+3}')^2-4\omega'^2-4i\omega'\Gamma_{2l+3}'
\over\Bigl[(2l+3+\Delta\omega_{2l+3}')^2-4\omega'^2+\Gamma_{2l+3}'^2\Bigr]^2
+16\omega'^2\Gamma_{2l+3}'^2}\Biggl\{\Bigl(2l+3+\Delta\omega_{2l+3}'
\nonumber\\& &+2\omega'-i\Gamma_{2l+3}'\Bigr) \Biggl(\Bigl(e^{-\beta
E_l}-e^{-\beta E_{l+1}}\Bigr){l+1+\Delta\omega_{l+1}'-\omega'-i\Gamma_{l+1}'\over
(l+1+\Delta\omega_{l+1}'-\omega')^2+\Gamma_{l+1}'^2}
\nonumber\\&
&-\Bigl(e^{-\beta E_{l+1}}-e^{-\beta E_{l+2}}\Bigr){(l+2+\Delta
\omega_{l+2}'+\omega'-i\Gamma_{l+2}')\over
(l+2+\Delta\omega_{l+2}'+\omega')^2+\Gamma_{l+2}'^2}\Biggr)(1-\alpha)\nonumber\\&
&+\Bigl(2l+3+\Delta\omega_{2l+3}'-2\omega'+i\Gamma_{2l+3}'\Bigr)
\Biggl(\Bigl(e^{-\beta E_{l}}-e^{-\beta
E_{l+1}}\Bigr)\nonumber\\&
&\times{l+1+\Delta\omega_{l+1}'+\omega'+i\Gamma_{l+1}')\over
(l+1+\Delta\omega_{l+1}'+\omega')^2+\Gamma_{l+1}'^2}-\Bigl(e^{-\beta
E_{l+1}}-e^{-\beta E_{l+2}}\Bigr)\nonumber\\& &\times{l+2+\Delta
\omega_{l+2}'+\omega'+i\Gamma_{l+2}'\over
(l+2+\Delta\omega_{l+2}'-\omega')^2+\Gamma_{l+2}'^2}\Biggr)(1-\alpha)\nonumber\\& &+\alpha(2l+3+\Delta\omega_{2l+3}') \Bigl(e^{-\beta E_l}-e^{-\beta
E_{l+2}}\Bigr)\Biggr\}\Biggr].     \end{eqnarray}Relation (\ref{eq:keb}) shows
how frequency-time dependent Kerr optical function depends on
field parameters like frequency and field strengths, on
the molecular parameters like moment of inertia, dipole moment and
polarisability, on the bath frictional parameter and on the temperature.
Despite the fact that these results have been obtained in the limit of small
coupling parameter $B$, they can still be used as first approximation to
interpret experimental results on dense bath but at very low temperatures. We
point out that, unlike earlier works on electrical susceptibility which have
always considered that observed spectra are mainly accounted for by transitions
involving $\Delta l=\pm 1$, these results on the Kerr optical effect predict,
not only $\Delta l=\pm 1$ transitions, but also those with $\Delta l=\pm 2.$ 

\section{Discussions}1) In constant temperature conditions, the response of a
dielectric  material to a low frequency  external a.c. field is strong and is in phase with the latter for low frictional oscillator-bath and /or large inertia molecules
 ($\gamma=0.5 $ in Figure 1). From the energetic point of view, this in-phase
aspect favours the external field effect on the rotator-bath system and thus
increases the system's ability to capture energy from the surrounding. On the
other hand, a totally different phenomenon is observed for high frequencies where a weak  response sets in, tending to annihilate the field effect by appearing in anti-phase with the it (Figure 2). The first convergent of the classical susceptibility function corresponding
to the Rocard result (the first  approximation of the inertia effect) gives a maximum phase for a frequency of $\omega=\sqrt{2}\omega_{mean}$ whatever the
$\gamma$ value. At this frequency value collisions result to large energy
exchanges of the order of $k_BT$. For higher $\gamma$ values, there is a
departure from this frequency value (see Figure 4).  
\newline 2) The manifestation of quantum effects depends, not
only on the coupling parameter ($B=\zeta/I$) but also on the temperature-inertia parameter
$a=\hbar^2/(Ik_BT)$. This allows us to define a necessary
condition for the domination of quantum effects. The inequality
\begin{equation}
s=B/\omega_{mean}=1/\sqrt{\gamma}<<{1\over 2}\Bigl(\hbar^2/Ik_BT\Bigr)^{3/2}=a^{3/2}/2
\end{equation}expresses this condition. For example, for $a=0.05$,
as the parameter $s$ decreases from $0.025$ through $0.010$ to $0.001$, we observe a passage from a continuous
classical spectrum through broadened lines to well defined discrete lines
(Figure 5) meanwhile for $a=.5$ quantum effects are already
present even for $s=.08$ (see Figure 6). This observation is also
important for the Kerr spectra (see Figure 14).
\newline 3)
 The time variation of the classical Kerr electrical birefringence (KEB) is characterised by oscillations
about $r-\omega$ dependent time constant values which decrease with increasing
$r$ and with increasing frequency. The $2\omega$-harmonic component is dominant
for small $r$ and large $\alpha$ values while the single $\omega$ one dominates for intermediate and higher $r$ values, (see Figures 7-9). Physically, the doubling
of period takes place by a process of progressive crushing of intermediate peak
values in the KEB curve shape with increasing $r$. This period change is noticed
by a set of pronounced transitions of non sinusoidal periodic regimes which takes
place between two sinusoidal regime limits corresponding to the extreme $r$
values ($E_a<<E_c$ and $E_a>>E_c$). 
\newline
4) For constant bath parameters and for small and intermediate $E_c/E_a$ ratio, the Kerr effect increases with increasing $\alpha$, presenting small amplitude distortions that disappear to form secondary peaks as $\alpha$ grows, portraying the progressive appearance of the influence of the $2\omega$-harmonic component. 
\newline
5) The Kerr spectral
function for a.c.-d.c. coupling is
 \begin{eqnarray}
\tilde{\Phi}(\omega,\Omega)&=&\Bigl[2{\it{Re}}\Phi_0(\omega)\delta(\Omega)+
\Phi_1(\omega)\delta(\Omega-\omega)+\Phi_1^*(\omega)\delta(\Omega+
\omega)\nonumber\\& &+\Phi_2(\omega)\delta(\Omega-2\omega)+
\Phi_2^*(\omega)\delta(\Omega+2\omega)\Bigr].
\end{eqnarray}This shows that for an a.c. field of given frequency, all three terms cannot be measured simultaneously.
The a.c.-d.c. field coupling on dielectrics, therefore, proves to be very useful as, depending on the harmonic component observed, we are able to predict the 
relative strengths of permanent dipole to induced dipole effects. The
$2\omega$-component dominates for polarisable  fluids
(large $\alpha$) while the $\omega$-one dominates for less polarisable fluids (small $\alpha$). More importantly, the observation of $2\omega$-harmonic
component may also entail that the most probable rotational lines involve
transitions like $l\rightarrow l\pm 2$ while the observation of $\omega$
harmonics concerns transitions $l\rightarrow l\pm 1$. This last point is very
important when $E_c$ and $E_a$ are of the same order of magnitudes.

\newpage
\section*{Figure Captions} 
${\bf Figure 1}$ shows the plots, for $\nu=0.15$, of: 1) \rule[.4ex]{4mm}{.2mm} $\cos \nu \tau$, and the reduced susceptibility $\chi_r(\nu, \tau)$ versus the dimensionless time $\tau=\omega_{mean}t$ for; 2) -\makebox[.4mm]{}-\makebox[.4mm]{}- $\gamma=0.05$,  3)  \raisebox{.4ex}{{\tiny $.  .  .  .$}} $\gamma=0.5$. 
\\${\bf Figure 2}$ shows the plots, for $\nu=4.00$, of: 1) \rule[.4ex]{4mm}{.2mm} $\cos \nu \tau$ and the reduced susceptibility $\chi_r(\nu,\tau)$ versus the dimensionless time $\tau=\omega_{mean}t$ for; 2) -\makebox[.4mm]{}-\makebox[.4mm]{}- $\gamma=0.05$ and 3)  \raisebox{.4ex}{{\tiny $.  .  .  .$}}  $\gamma=0.5$ .
\\${\bf Figure 3}$ shows the 3D plot, for $\gamma=0.05 $ and $r=0$, of the reduced polarisation $\chi_r(\omega', \tau)$ versus the dimensionless time and freqency $\tau=Bt$ and $\omega'=\omega/B$, respectively.
\\${\bf Figure 4}$ shows the 3D plot, for $\gamma=0.05$, of $\tan(\alpha(\nu,\gamma))$ versus the dimensionless  freqency $\nu=\omega/\omega_{mean}$ and $\gamma$ (classical result).
\\${\bf Figure 5}$ shows the plots of $\alpha(\omega')$ (in $rad$) versus the dimensionless frequency $\omega'=\omega/(\hbar/I)$ for $a=0.05$: 1) \raisebox{.4ex}{{\tiny $.  .  .  .$}} $s=0.025$,
2) -\makebox[.4mm]{}-\makebox[.4mm]{}- $s= 0.01$ and 3) \rule[.4ex]{4mm}{.2mm} $s=0.001$ (RWA).
\\${\bf Figure 6}$ shows the plots of $\alpha(\omega')$ (in $rad$) versus the dimensionless frequency $\omega'=\omega/(\hbar/I)$ for $a=0.5$: 1) \raisebox{.4ex}{{\tiny $.  .  .  .$}} $s=0.25$, 2)  -\makebox[.4mm]{}-\makebox[.4mm]{}- $s= 0.08$ and 3)  \rule[.4ex]{4mm}{.2mm} $s=0.01$ (RWA).
\\$\bf{Figure 7}$ shows the  plots of the classical Kerr function versus the
dimensionless time $ t'=Bt$ for $\omega/B=1.0$,
$\alpha=0.1$: 1) \rule[.4ex]{4mm}{.2mm}  $r=0.95$:  2) -\makebox[.4mm]{}-\makebox[.4mm]{}- $r=0.5$ and  3) \raisebox{.4ex}{{\tiny $.  .  .  .$}} $r=0.05$.
\\$\bf{Figure 8}$ shows the  plots of the classical Kerr function versus the
dimensionless time $ t'=Bt$ for $\omega/B=1.0$,
$\alpha=0.5$: 1) \rule[.4ex]{4mm}{.2mm} $r=0.95$:  2)  -\makebox[.4mm]{}-\makebox[.4mm]{}- $r=0.5$ and  3) \raisebox{.4ex}{{\tiny $.  .  .  .  .$}} $r=0.05$.
\\$\bf{Figure 9}$ shows the  plots of the classical Kerr function versus the
dimensionless time $ t'=Bt$ for $\omega/B=1.0$,
$\alpha=0.9$: 1) \rule[.4ex]{4mm}{.2mm}   $r=0.95$: 2) -\makebox[.4mm]{}-\makebox[.4mm]{}- $r=0.5$ and  3) \raisebox{.4ex}{{\tiny $.  .  .  .$}} $r=0.05$.
\\$\bf{Figure 10}$ shows the  plots of the quantum Kerr function versus the
dimensionless time $ t'=t/(\hbar/I)$ for $\omega=4\hbar/I$,
$\alpha=0.1$: 1)  \rule[.4ex]{4mm}{.2mm} $r=0.95$:  2)   -\makebox[.4mm]{}-\makebox[.4mm]{}- $r=0.5$ and  3) \raisebox{.4ex}{{\tiny $.  .  .  .$}} $r=0.05$.
\\$\bf{Figure 11}$ shows the  plots of the quantum Kerr function versus the
dimensionless time $ t'=t/(\hbar/I)$ for $\omega=4\hbar/I$,
$\alpha=0.5$: 1) \rule[.4ex]{4mm}{.2mm} $r=0.95$:  2) -\makebox[.4mm]{}-\makebox[.4mm]{}-  $r=0.5$ and  3)\raisebox{.4ex}{{\tiny $.  .  .  .$}}  $r=0.05$.
\\$\bf{Figure 12}$ shows the  plots of the quantum Kerr function versus the
dimensionless time $t'=t/(\hbar/I) $ for $\omega=4\hbar/I $,
$\alpha=0.9$: 1) \rule[.4ex]{4mm}{.2mm} $r=0.95$:  2) -\makebox[.4mm]{}-\makebox[.4mm]{}-   $r=0.5$ and  3) \raisebox{.4ex}{{\tiny $.  .  .  .$}}
$r=0.05$.
\\$\bf{Figure 13}$ shows the  plot of the
real part of   1)  -\makebox[.4mm]{}-\makebox[.4mm]{}- the  $\omega$-harmonic component and 2) \rule[.4ex]{4mm}{.2mm} the $2\omega$-harmonic component of the quantum Kerr function versus the dimensionless frequency $\omega/(\hbar/I)$ for $a=0.05$,   $s=0.01$, $\alpha=0.4$ and $r=0.5$.
\\$\bf{Figure 14}$ shows the  plot of the imaginary part of the $\omega$-harmonic component of the quantum Kerr function versus the dimensionless frequency $\omega/(\hbar/I)$ for 1)   \rule[.4ex]{4mm}{.2mm} $a=0.25$, $s=0.10$, $\alpha=0.4$, $r=0.5$ and 2)  -\makebox[.4mm]{}-\makebox[.4mm]{}- $a=0.05$, $s=0.01$, $\alpha=0.4$, $r=0.5$. 

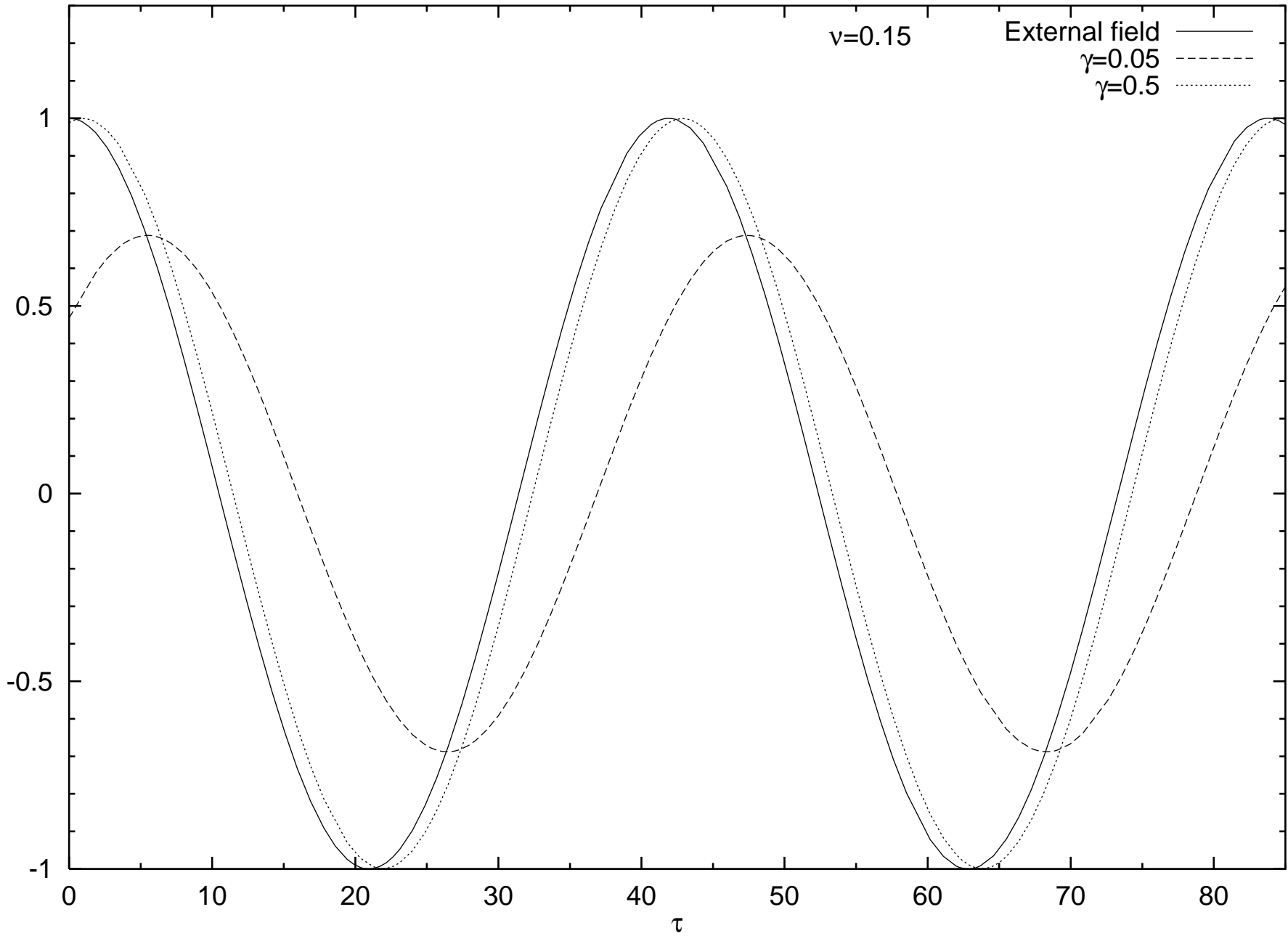
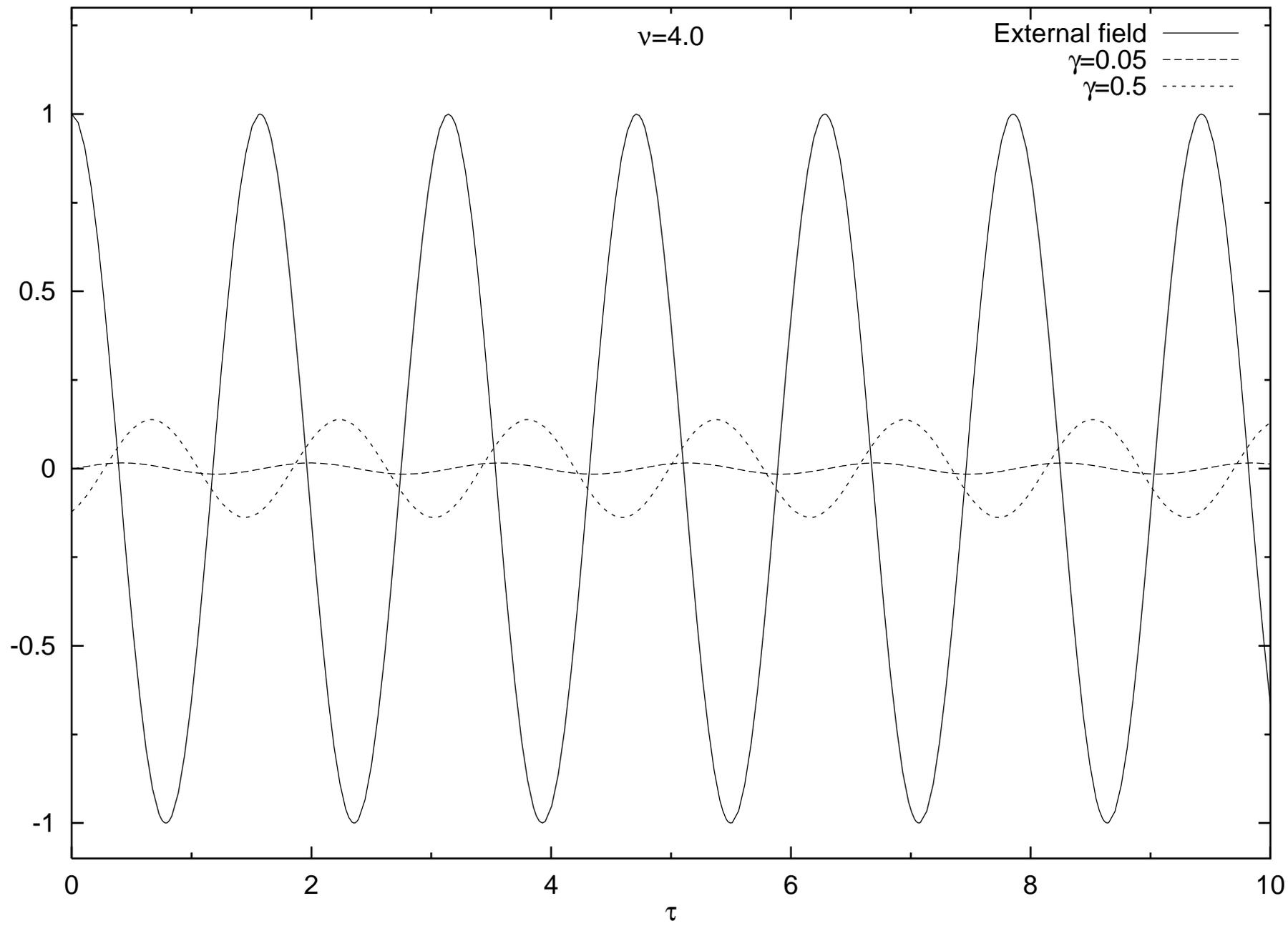
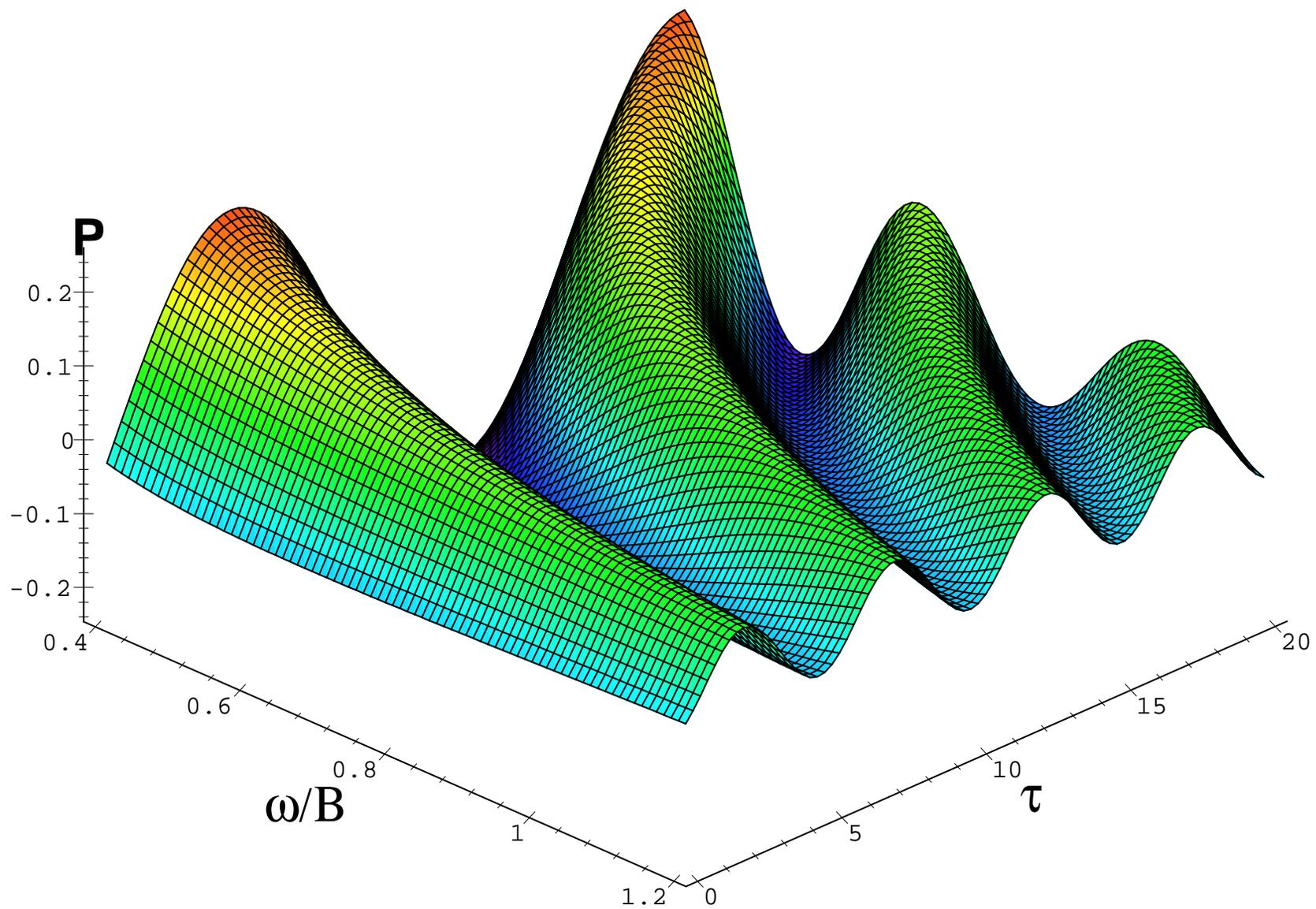
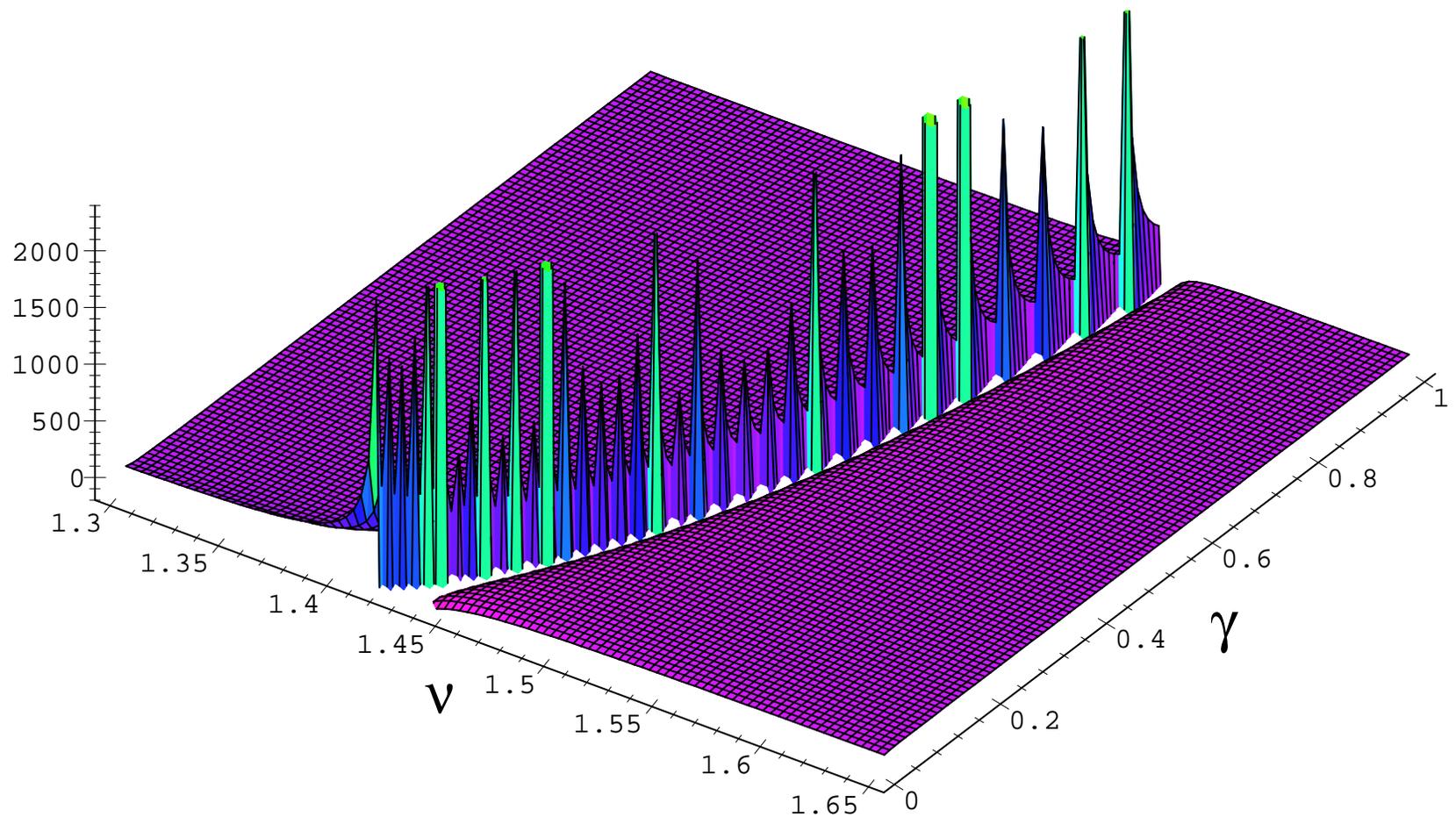
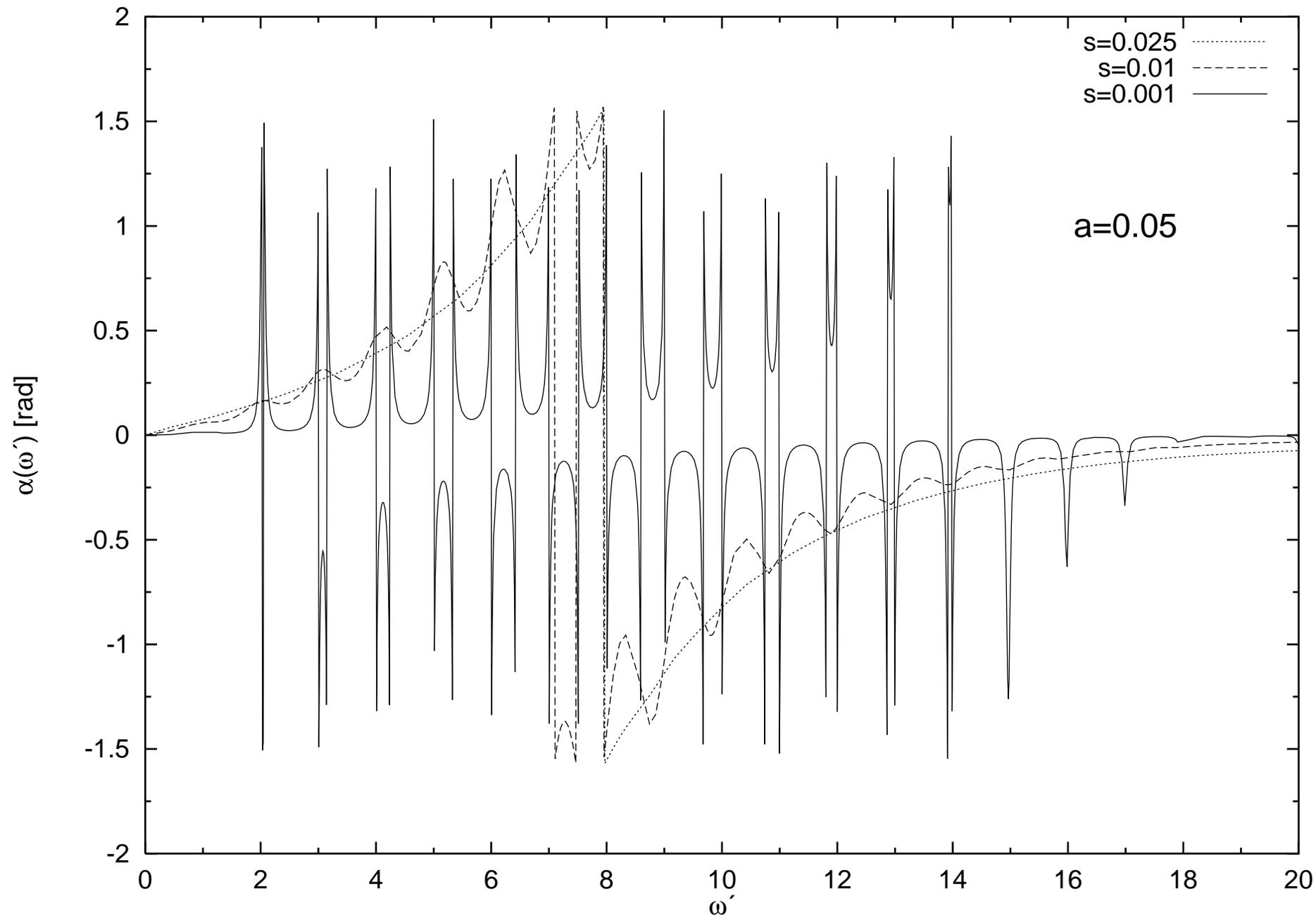
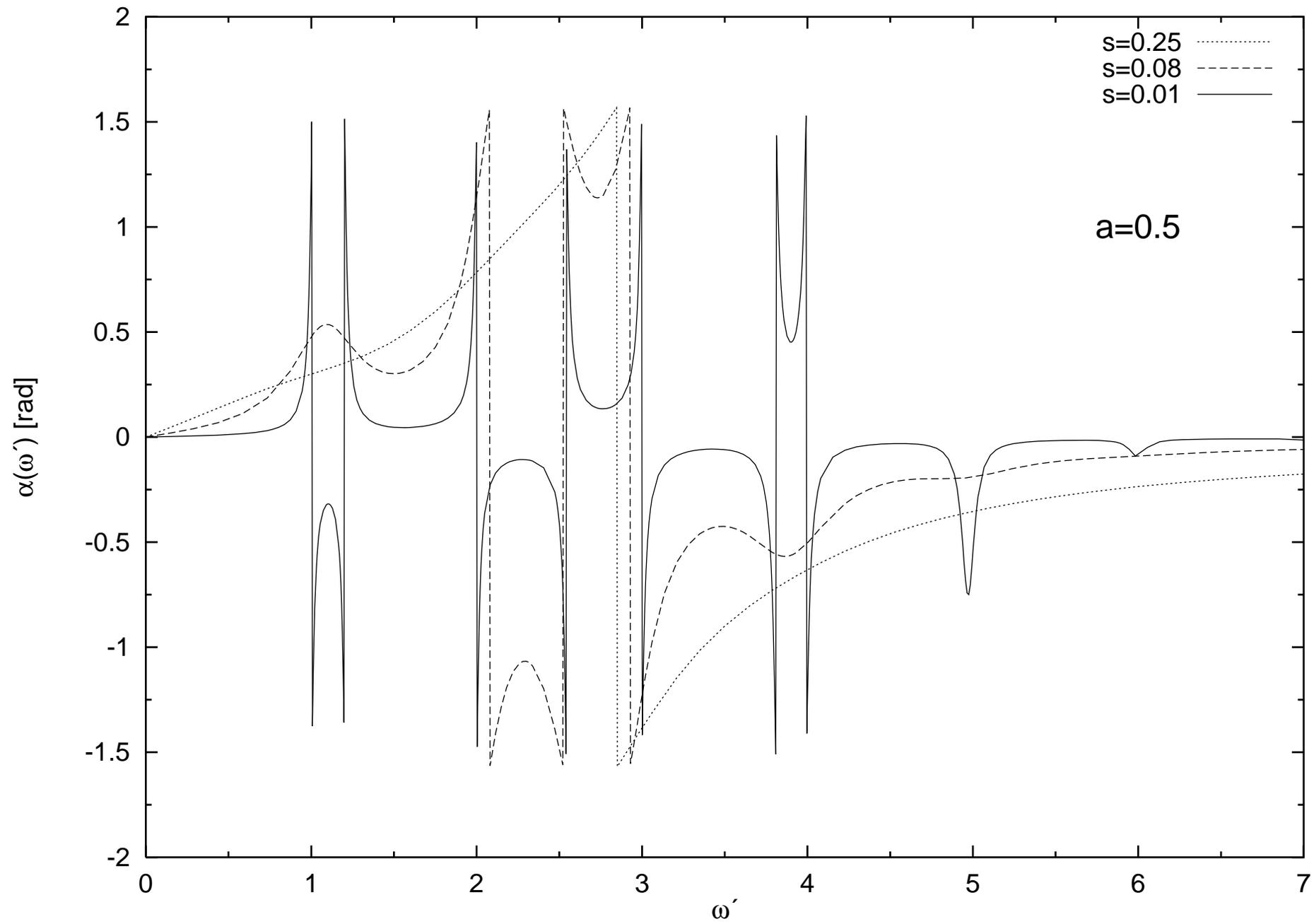
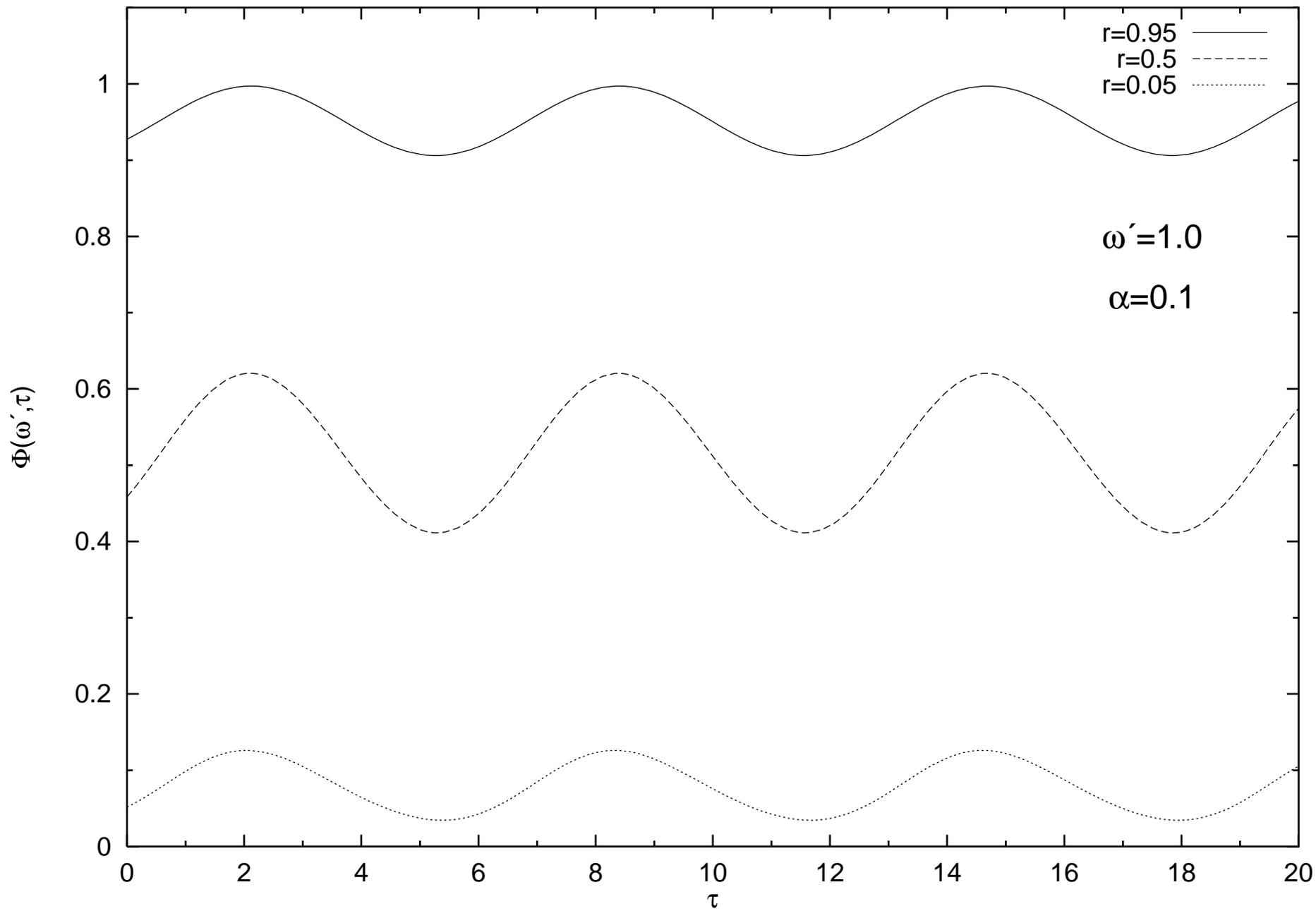
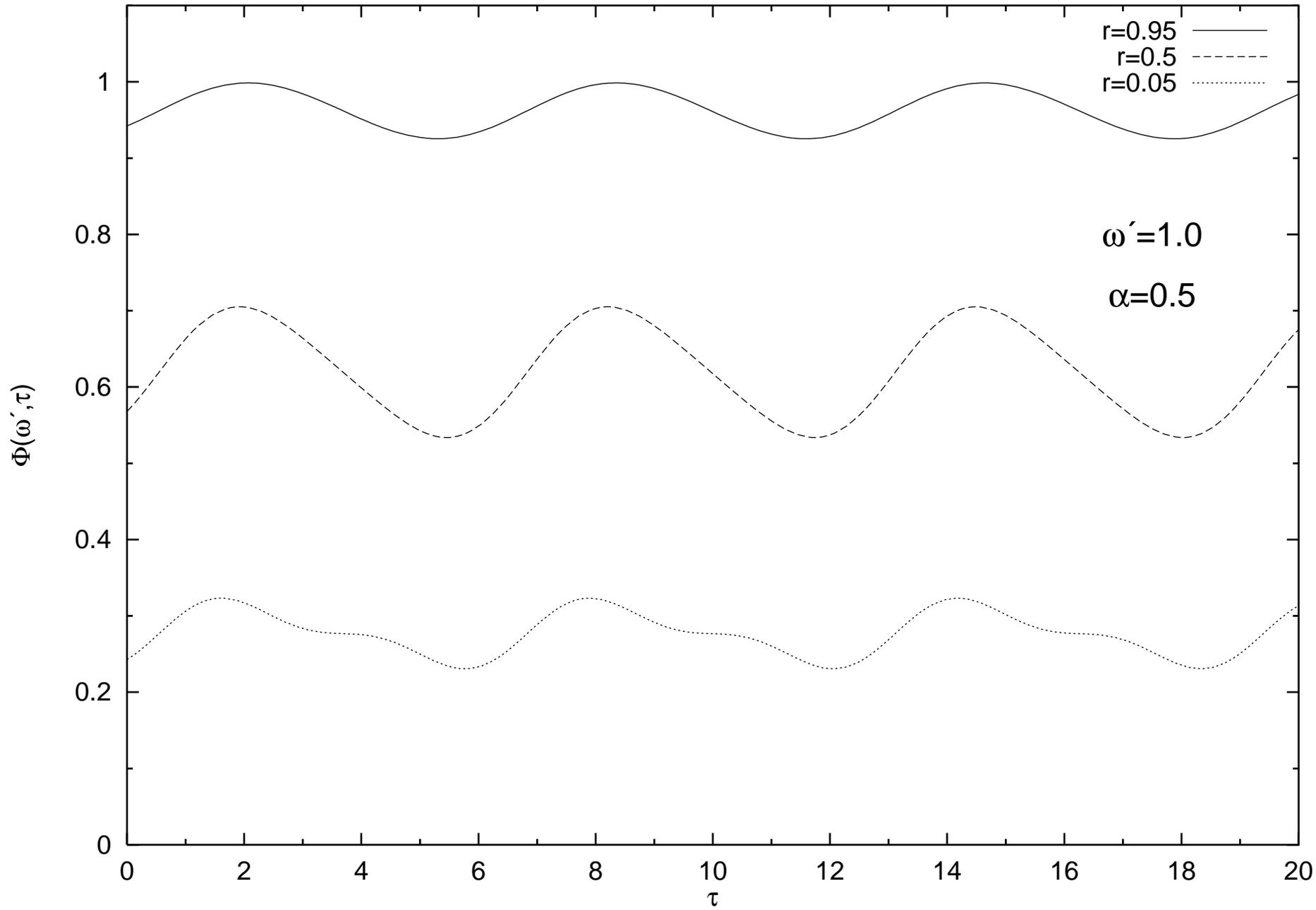
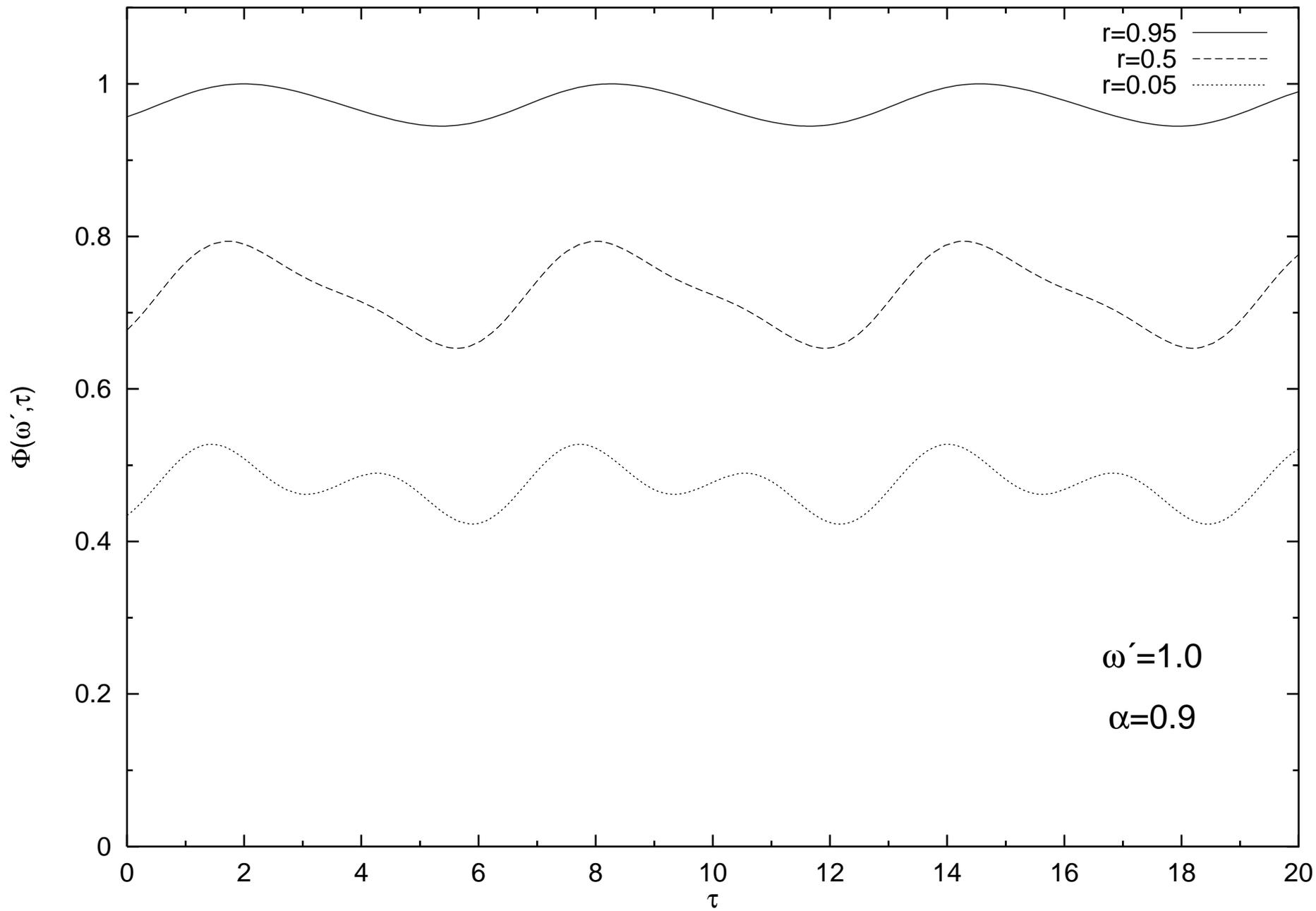
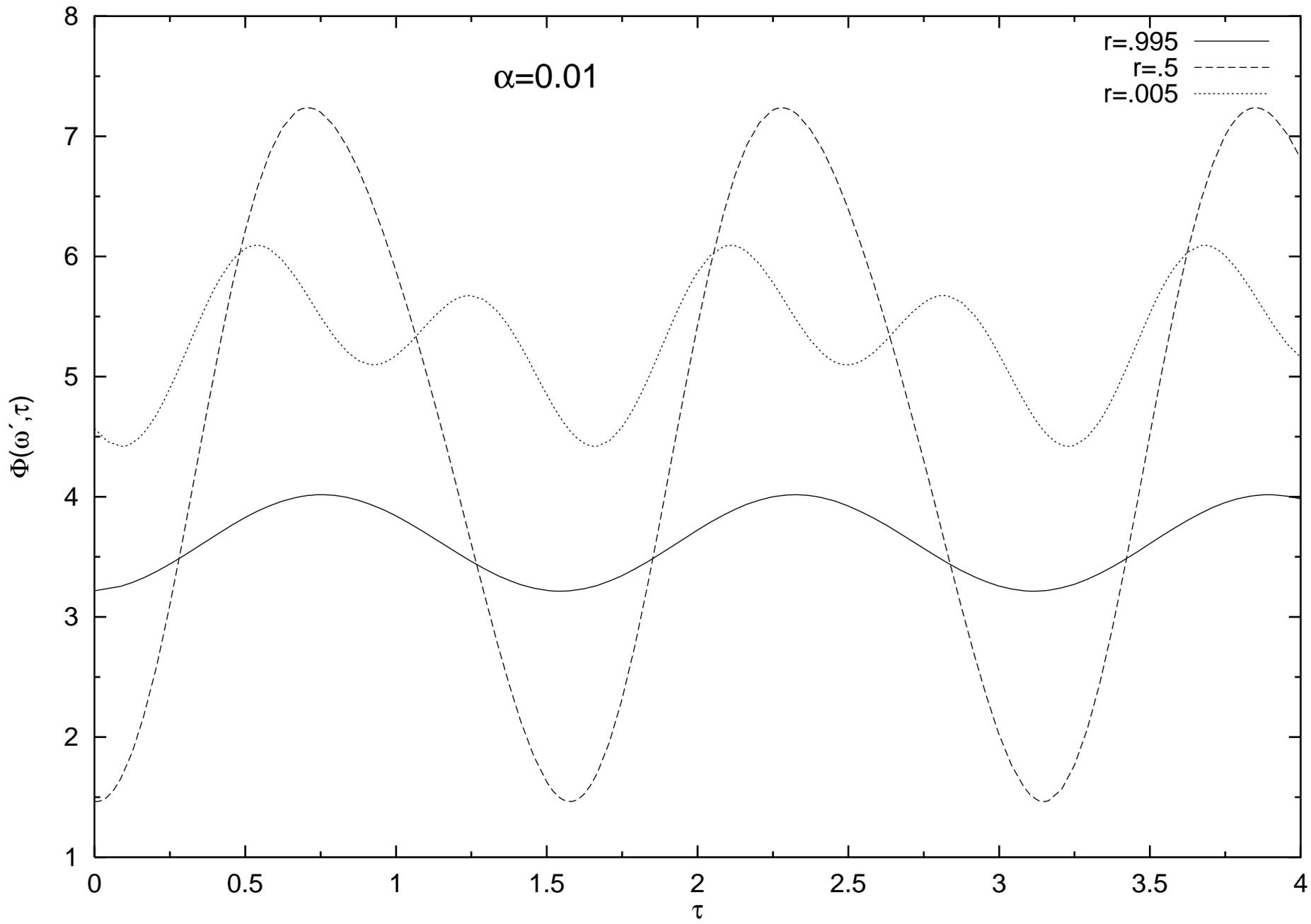
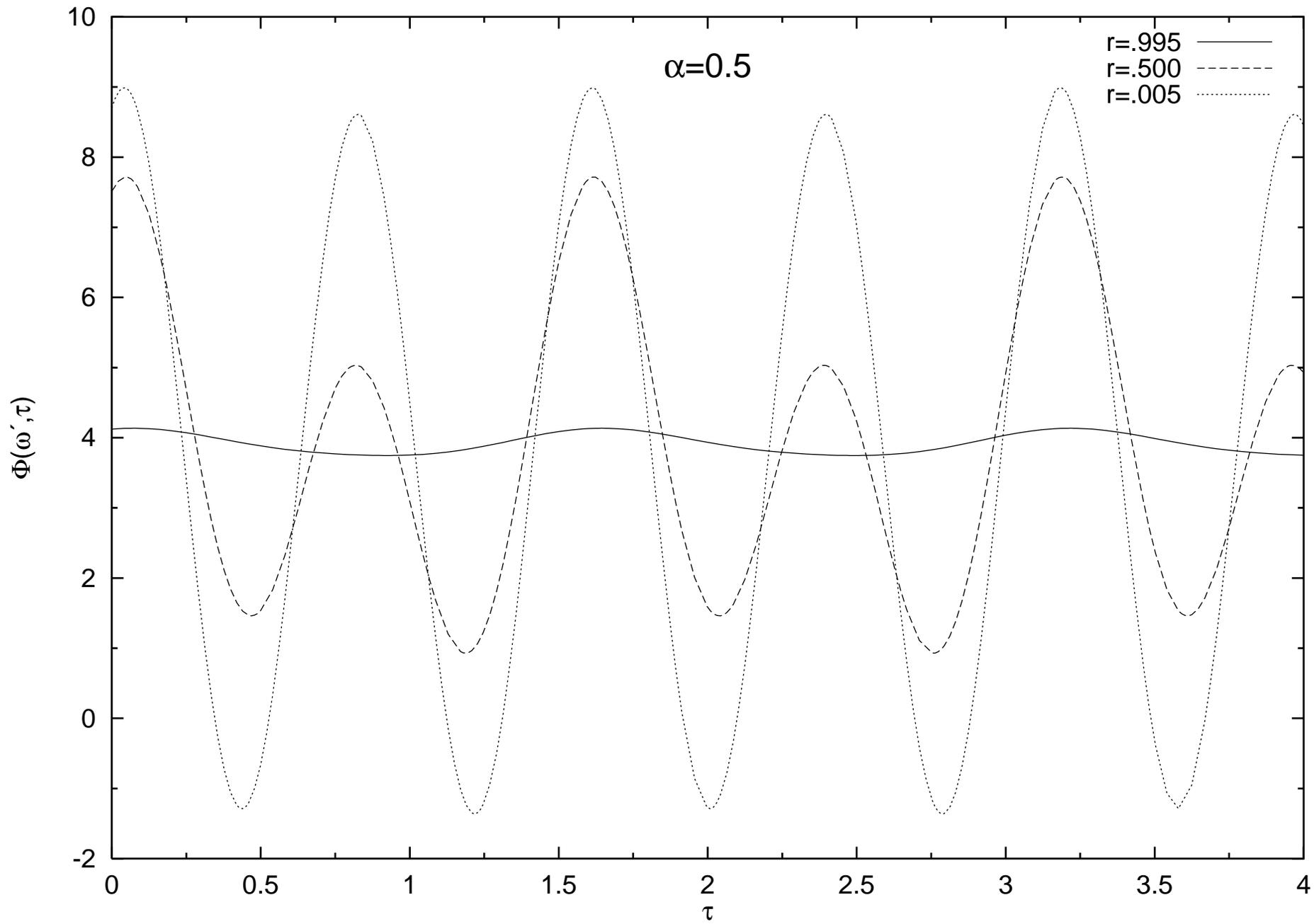
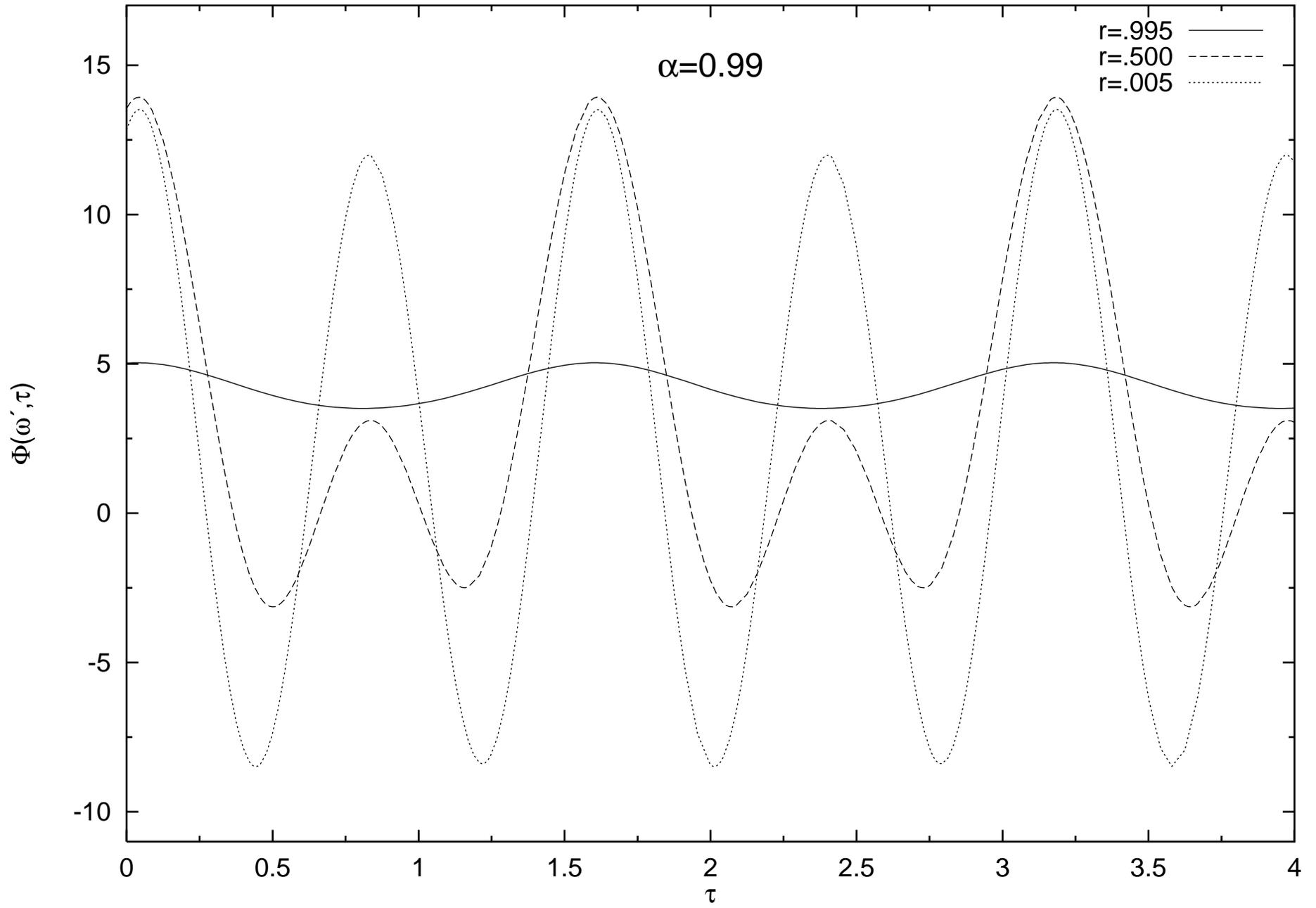
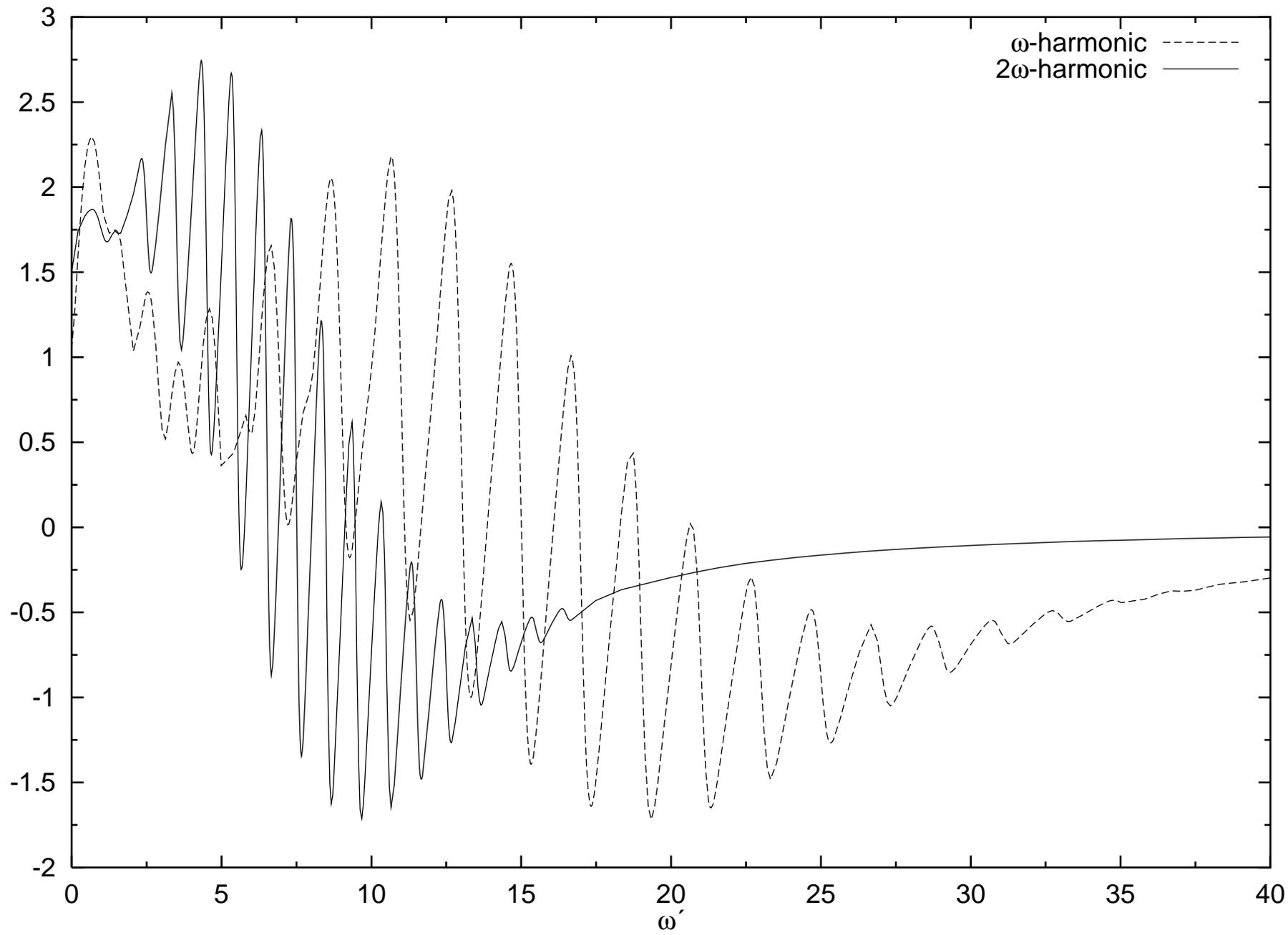
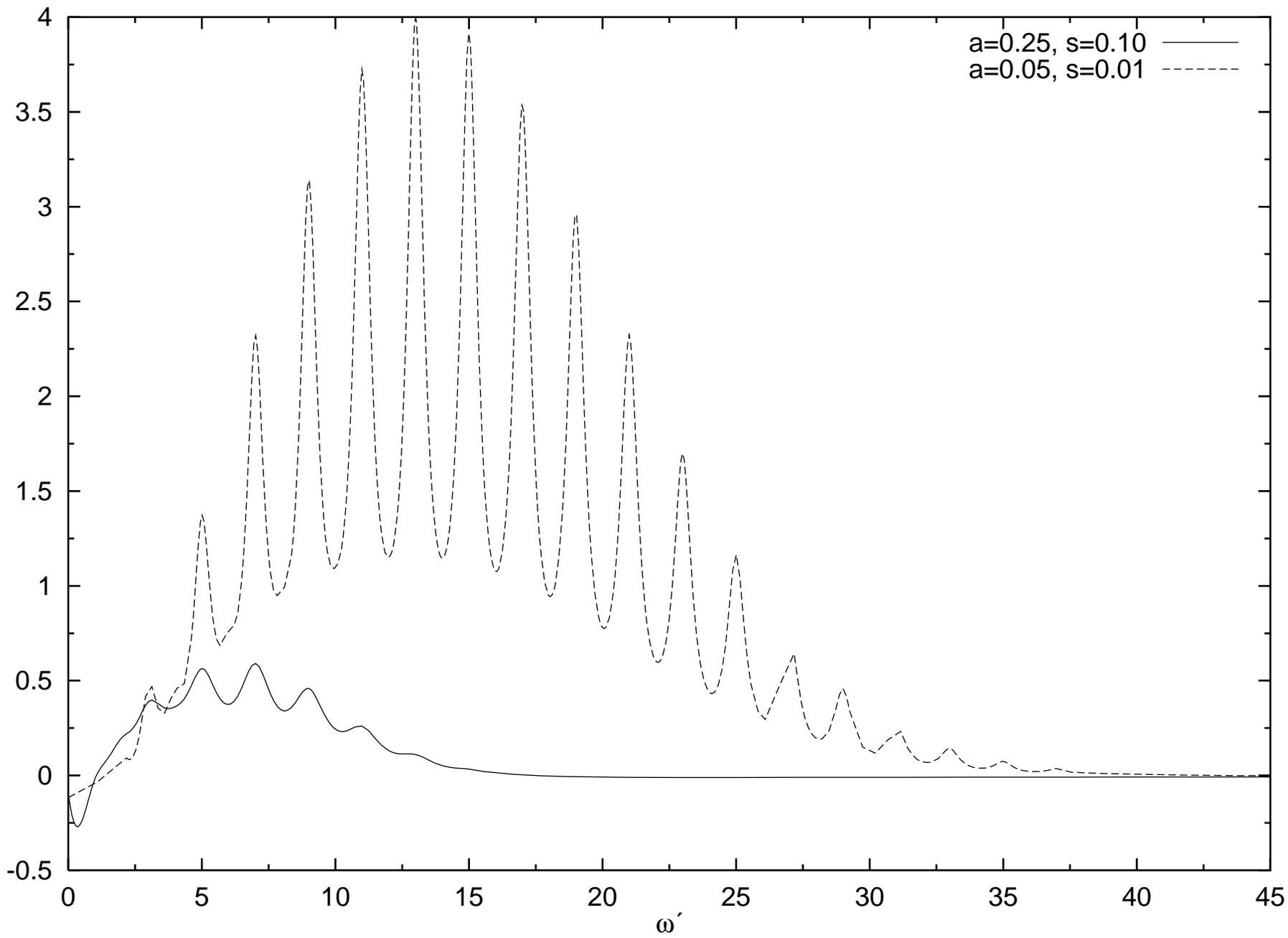
\end{document}